\documentclass[
  notitlepage,
  twocolumn,
  preprintnumbers,
  longbibliography,
  superscriptaddress,
  aps,
  prl,
  amsmath,
  amssymb,nofootinbib
]{revtex4-2}

\usepackage{graphicx,subfigure,epsfig}
\usepackage{dcolumn}
\usepackage{amssymb}
\usepackage{times}
\usepackage{amsmath}
\usepackage{amsfonts}
\usepackage{mathrsfs}
\usepackage{setspace}
\usepackage{latexsym}
\usepackage{bbm}
\usepackage{float}
\usepackage{flafter}
\usepackage{bm}
\usepackage{epstopdf}
\usepackage[dvipsnames]{xcolor}
\usepackage{multirow}
\usepackage{physics} %Dirac bracket
\usepackage[export]{adjustbox}
\DeclareMathOperator{\sgn}{sgn}
\usepackage{braket}
\usepackage{mathtools}
\usepackage{dsfont}
\usepackage[normalem]{ulem}

\usepackage{soul} % for underline with automatic line-break

\def \be {\begin{eqnarray}}
\def \ee {\end{eqnarray}}

\newcommand{\bk}{{\bf k}}

\newcommand{\bq}{{\bf q}}
\newcommand{\bx}{{\bf x}}
\newcommand{\by}{{\bf y}}
\newcommand{\bz}{{\bf z}}

\newcommand{\bee}{{\bf e}}

\newcommand{\bG}{{\bf G}}

\newcommand{\br}{{\bf r}}
\newcommand{\bR}{{\bf R}}
\newcommand{\bs}{{\bf s}}

\newcommand{\bn}{{\bf n}}

\newcommand{\bt}{{\bf t}}
\newcommand{\bv}{{\bf v}}

\usepackage{dsfont}

\usepackage{xcolor}

 \usepackage{simpler-wick} % for wicks contraction

\newcommand{\bea}{\begin{equation} \begin{aligned}}
\newcommand{\eea}{\end{aligned} \end{equation} }

\newcommand{\bpm}{\begin{pmatrix}}
\newcommand{\epm}{\end{pmatrix}}
\newcommand{\eps}{\epsilon}

\renewcommand{\Tr}{\text{Tr}}

\newcommand{\cc}[1]{\langle{#1}\rangle_c}

\usepackage{xr}
\usepackage[pdftex,breaklinks,bookmarks=false,colorlinks,linkcolor=blue,citecolor=blue,urlcolor=blue,hypertexnames=false]{hyperref}

\usepackage{cleveref}
\crefname{appendix}{App.}{Apps.}
\crefname{equation}{Eq.}{Eqs.}
\crefname{figure}{Fig.}{Figs.}
\crefname{table}{Tab.}{Tabs.}
\crefname{section}{Sec.}{Secs.}
\creflabelformat{appendix}{#2#1#3}

\usepackage{wrapfig}

\usepackage{orcidlink}

\usepackage[symbol]{footmisc}

\def\@fnsymbol#1{\ensuremath{\ifcase#1\or *\or \dagger\or \ddagger\or
   \mathsection\or ** \or\mathparagraph\or \|\or \dagger\dagger
   \or \ddagger\ddagger \else\@ctrerr\fi}}

\begin{document}

\title{Fermi Surface Geometry from Charge Fluctuations in Three-Dimensional Metals}

\author{Pok Man Tam}
\email{pmtamaa@gmail.com}
\thanks{These two authors contributed equally.}
\affiliation{Princeton Center for Theoretical Science, Princeton University, Princeton, New Jersey 08544, USA}

\author{Yarden Sheffer}
\email{yarden.sheffer@gmail.com}
\thanks{These two authors contributed equally.}
\affiliation{Department of Condensed Matter Physics, Weizmann Institute of Science Rehovot 7610001, Israel}

\author{Xiao-Chuan Wu}
\email{xiaochuanwu.physics@gmail.com}
\affiliation{Department of Physics, Princeton University, Princeton, New Jersey 08544, USA}

\author{F. D. M. Haldane}
\email{haldane@princeton.edu}
\affiliation{Department of Physics, Princeton University, Princeton, New Jersey 08544, USA}

\author{Shinsei Ryu}
\email{shinseir@princeton.edu}
\affiliation{Department of Physics, Princeton University, Princeton, New Jersey 08544, USA}

\date{\today}

\begin{abstract}
For three-dimensional non-interacting multi-band metals, we show that important information about the \textit{shape} and the \textit{quantum geometry} of Fermi surfaces is encoded in the subleading logarithmic term of bipartite charge fluctuations. This logarithmic term is related to the dimensionless $\abs{\bq}^3$-coefficient of the structure factor in momentum space, and both quantities can be expressed as Fermi surface integrals of the Fermi surface curvature tensor and the quantum metric tensor. When the real-space partition surface is a quadric (i.e., sphere or ellipsoid), the logarithmic coefficient satisfies a topological bound depending only on the Euler characteristic and the Chern number of the Fermi surface, illustrating a non-trivial interplay between topology and quantum topology in multi-band metals.
\end{abstract}

\maketitle

\noindent{\color{blue}\emph{Introduction--}} 
Geometry and topology provide powerful frameworks for characterizing many-body quantum matter \cite{provost1980riemannian, berry1984quantal, shapere1989geometric, SWM2000, HasanKane2010, Resta2011Rev, bengtsson2017geometry, moessner2021topological, Torma2023essay, yu2025quantum}. In metals, the most common gapless quantum matter, their ground states are characterized by Fermi surfaces in the momentum space, whose \textit{shape} can be characterized both in terms of geometry and topology \cite{kaganov1979electron,Lifshitz1960,Haldane1994,Kane2022a,TamKane2022a,Yang2022, TamKane2022b,Zhang2023,TamKane2023a, Yang2023,TamKane2023b,Jia2025,daix2025probing,tam2026singular}. Moreover, as there is a manifold of quantum states defined on the Fermi surface, there is a notion of \textit{quantum geometry} associated to the variation of wavefunctions on the Fermi surface \cite{Haldane2004, NagaosaRMP2010,haldane2014attachment,chen2017berry, Alexandradinata2018, Dai2020,alexandradinata2023fermiology,yu2024non}. While geometric and topological aspects of insulators have been extensively explored, much less is known for gapless metals. Here we address a central question about how these mathematical characterizations manifest in physical observables through charge fluctuations in non-interacting metals.

Charge fluctuations can be quantified using the \textit{equal-time} connected density-density correlation function, $S(\bq)$, also known as the structure factor, which can be organized in powers of momentum as 
\begin{equation}\label{eq: expansion of Sq}
	S(\bq)  = S^{(1)}(\bq) +S^{(2)}(\bq) -S^{(3)}(\bq) +\mathcal{O}(\abs{\bq}^4),
\end{equation}
with $S^{(d)}(\bq) = S^{(d)}(\hat{\bq})\abs{\bq}^d$. Here $S(\bq) \equiv \cc{\rho_\bq \rho_{-\bq}}/{V}$, $\rho_\bq$ is the fermion density at momentum $\bq$ ($\hat{\bq}\equiv\bq/\abs{\bq}$) and $V$ is the system volume. Notably, in $D$ spatial dimensions, the coefficient of $\abs{\bq}^D$ is a \textit{dimensionless} quantity that can encode fundamental information about the many-body system. The $D=1$ case is well-known: for 1D free Fermi gases, $S^{(1)}$ is topological and counts the number of gapless Fermi points. In a single-channel Luttinger liquid, $S^{(1)}$ is no longer quantized but reflects the Luttinger parameter encoding interaction effects \cite{Giamarchi2004}. For $D=2$, recent works have connected $S^{(2)}$ to the quantum geometry in insulators \cite{Regnault2013, Onishi2025, Tam_corner2024, Wu_corner2025}, through the integrated quantum metric that characterizes localization \cite{Kivelson1982, MarzariVanderbilt1997, Resta1999, SWM2000,Ryu2010MomentumSpaceMetric}. Naturally, one would ask what is encoded in $S^{(3)}$ in $D=3$. The non-analytic nature of $\abs{\bq}^3$ suggests its connection to long-range correlations in gapless systems. As we will explain, it captures interesting geometric as well as quantum geometric aspects of Fermi surfaces in three dimensions. 

The consideration of $S(\bq)$ is closely related to bipartite charge fluctuations in real space, $\cc{Q_A^2}$, where $Q_A$ measures the charge (i.e., particle number) in a subregion $A$ with a linear size $L$. We first focus on spherical partitions. For odd-$D$ dimensions, the singular $\abs{\bq}^D$ term in Eq.~\eqref{eq: expansion of Sq} is associated to a logarithmic term in the finite-size scaling of $\cc{Q^2_A}$. For $D=3$, which is the focus of this work, we have
\begin{equation}\label{eq: def of Gamma}
	\cc{Q^2_A} \supset -\Gamma \log L,\;\;\;\Gamma = \frac{1}{\pi}\int d^2\hat{\bq}\;S^{(3)}(\hat{\bq}),
\end{equation}
where $\supset$ indicates a specific finite-size scaling contributing to $\cc{Q^2_A}$. The $D=1$ case is again well-known, and by the cumulant expansion relating charge fluctuations to entanglement entropy in free fermions \cite{KlichLevitov2009, Song2011, Song2012, calabrese2012exact}, the logarithmic coefficient of entanglement entropy is associated to the central charge in a ($1+1$)-dimensional conformal field theory (CFT) \cite{holzhey1994geometric, calabrese2004entanglement, GioveKlich2006, FradkinMooire2006, Swingle2010, Swingle2012,SwingleSenthil2013, KunYang2012,TanRyu2020}. Interestingly, in a ($3+1$)-dimensional CFT, the logarithmic term in entanglement entropy is related to the conformal anomaly \cite{solodukhin2008entanglement, casini2010entanglement}. 
Here, instead of studying conformal invariant systems and entanglement entropy, we study charge fluctuations in gapless Fermi surface systems, which should be experimentally accessible via inelastic x-ray scattering in solid-state settings \cite{schulke2007electron, Abbamonte2025} or quantum gas microscopy in ultracold atom quantum simulators \cite{nelson2007imaging,3DQGM2020,Bakr_review}. We find that $S^{(3)}$ and $\Gamma$ can be expressed as integrals of Fermi surface
curvature and quantum metric over the Fermi surface. Furthermore, $\Gamma$ has a topological lower bound determined by the Fermi surface Euler characteristic and Chern number. The precise geometric content of $\Gamma$ is illustrated with two examples. We then generalize to ellipsoidal partitions and comment on interaction effects as well as implications for entanglement and anomalies.

\noindent{\color{blue}\emph{General consideration--}}
The electron density operator $\rho_\bq$ takes the following form in a generic multi-band system:
\begin{equation}\label{eq: eq: general form for density}
    \rho_\bq = \sum_{\bk\in\rm BZ}\sum_{\sigma, m, n} U^*_{\sigma,m}(\bk)U_{\sigma, n}(\bk+\bq) c^\dagger_{m,\bk} c_{n,\bk+\bq}
\end{equation}
where  $\sigma$ is the unit-cell orbital index, $m$ and $n$ are band indices, $U_{\sigma,m}(\bk)$ is the normalized $m$-th band Bloch eigenvector at momentum $\bk$, and $c_{m,\bk}$ is the corresponding annihilation operator. The momentum summation is restricted within the first Brillouin zone (BZ). The above expression is obtained from $\rho_\bq = \int d^3\br\rho(\br)e^{-i\bq\cdot\br} = \sum_{\bR,\sigma}e^{-i\bq\cdot\bR_\sigma} c^\dagger_{\bR,\sigma}c_{\bR,\sigma}$, where $c_{\bR,\sigma} = \frac{1}{\sqrt{N_c}}\sum_{\bk\in{\rm BZ}, m}e^{i\bk\cdot\bR_\sigma}U_{\sigma,m}(\bk)c_{m,\bk}$ is the annihilation operator for the $\sigma$-orbital in the unit cell (labeled by $\bR$), positioned at $\bR_\sigma$, and $N_c$ is the number of unit cells. 

The bipartite charge fluctuation, $\cc{Q^2_A} \equiv \langle Q^2_A\rangle - \langle Q_A\rangle^2$, can be expressed as
\begin{equation}
\begin{split}
    \cc{Q^2_A} &= \int\frac{d^3\bq}{(2\pi)^3}\int\frac{d^3\bq'}{(2\pi)^3} \mathcal{F}_A(\bq)\mathcal{F}_A(\bq')\cc{\rho_\bq\rho_{\bq'}}
\end{split}
\end{equation}
where $\mathcal{F}_A(\bq) \equiv \int_A d^3\br e^{i\bq\cdot\br}$ is the form factor for the spatial partition region $A$, and $\mathcal{F}_A(\bq) = \frac{4\pi}{\abs{\bq}}[\sin(\abs{\bq}L)-\abs{\bq}L\cos(\abs{\bq}L)]$ when the partition is spherical of radius $L$. Using Eq.~\eqref{eq: eq: general form for density}, it can be observed that $\cc{\rho_\bq\rho_{\bq'}}$ is non-vanishing only when $\bq' = -\bq+\bG$ for some reciprocal lattice vector $\bG$. Furthermore, \textit{logarithmic} finite-size scaling can be shown to arise only from having $\bq'=-\bq$ (see supplemental material (SM) \cite{supp}), corresponding to $\int \frac{d^3\bq}{(2\pi)^3}\abs{\mathcal{F}_A(\bq)}^2S(\bq)$. Combining this with Eq.~\eqref{eq: expansion of Sq}, Eq.~\eqref{eq: def of Gamma} follows.

We now examine $S(\bq)$ in detail. In the thermodynamic limit and at zero temperature, following Wick's theorem, 
\begin{equation}\label{eq: multiband Sq}
    S(\bq)= \int_{\rm BZ} \frac{d^3\bk}{(2\pi)^3}\sum_{m,n} \mathscr{P}_{m,n}(\bk, \bq) f_{m,\bk}(1-f_{n,\bk+\bq}).
\end{equation}
Here $f_{m,\bk}=\theta(E_F-E_{m,\bk})$ is the Fermi occupation for the $m$-th band, where $E_{m,\bk}$ is the band dispersion and $E_F$ is the Fermi energy. Quantum geometry is encoded in
$\mathscr{P}_{m,n}(\bk, \bq)\equiv\Tr[P_{m}(\bk) P_{n}(\bk+\bq)]$, where $[P_{m}(\bk)]_{\sigma,\sigma'} \equiv U_{\sigma, m}(\bk)U^*_{\sigma',m}(\bk)$ is the $m$-th band projector, with $\Tr$ tracing over the orbital space. Expanding $\mathscr{P}_{m,n}(\bk, \bq)$ in powers of $q$,  
$S(\bq)$ is decomposed into two parts: the \textit{geometric} part (denoted by ${\rm G}$) depends only on the shape of the Fermi surface and arises from the $q^0$-term of $\mathscr{P}_{m,n}(\bk, \bq)$ (i.e., $\delta_{m,n}$); the \textit{quantum geometric} part (denoted by ${\rm QG}$) depends on the band-projector and arises from $q^{>0}$ terms of $\mathscr{P}_{m,n}(\bk, \bq)$. Correspondingly, 
\begin{equation}
    S^{(3)}(\bq) = S^{(3)}_{\rm G}(\bq)+S^{(3)}_{\rm QG}(\bq),\quad\Gamma=\Gamma_{\rm G}+\Gamma_{\rm QG}.
\end{equation}
These are the central quantities of our interest. We first focus on the geometric part.

\noindent{\color{blue}\emph{Spherical Fermi surfaces--}}
Insights can be acquired by analyzing the simplest case in a Fermi gas with a spherical Fermi surface. With a single band,
\begin{equation}\label{eq: single band Sq}
    S(\bq) = \int_{\rm BZ} \frac{d^3\bk}{(2\pi)^3}\left(f_\bk-f_\bk f_{\bk+\bq}\right),
\end{equation}
which can be interpreted geometrically as the difference between the volume of the Fermi sea and the \textit{overlapping volume} between the original Fermi sea and its counterpart shifted by $\bq$. For a spherical Fermi surface with radius $k_F$, $S(\bq)= \abs{\bq}k_F^2/(8\pi^2) - \abs{\bq}^3/(96\pi^2)$ for $\abs{\bq}\leq 2k_F$. Notably, $S^{(3)}_{\rm G}(\hat{\bq})=1/(96\pi^2)$ is independent of $k_F$, which is expected from dimensional analysis. Similarly, for $N_F$ spherical Fermi surfaces which could independently vary in sizes and positions,
\begin{equation}\label{eq: gamma and Gamma for spheres}
	S^{(3)}_{\rm G}(\hat{\bq})=\frac{N_F}{96\pi^2} \quad\text{and}\quad \Gamma_{\rm G}=\frac{N_F}{24\pi^2}.
\end{equation}
This suggests a connection between charge fluctuations and the topology of Fermi surfaces, which we clarify below by analyzing Fermi surfaces with generic shapes.

\begin{figure}[t]
    \centering
    \resizebox{\columnwidth}{!}{\includegraphics[]{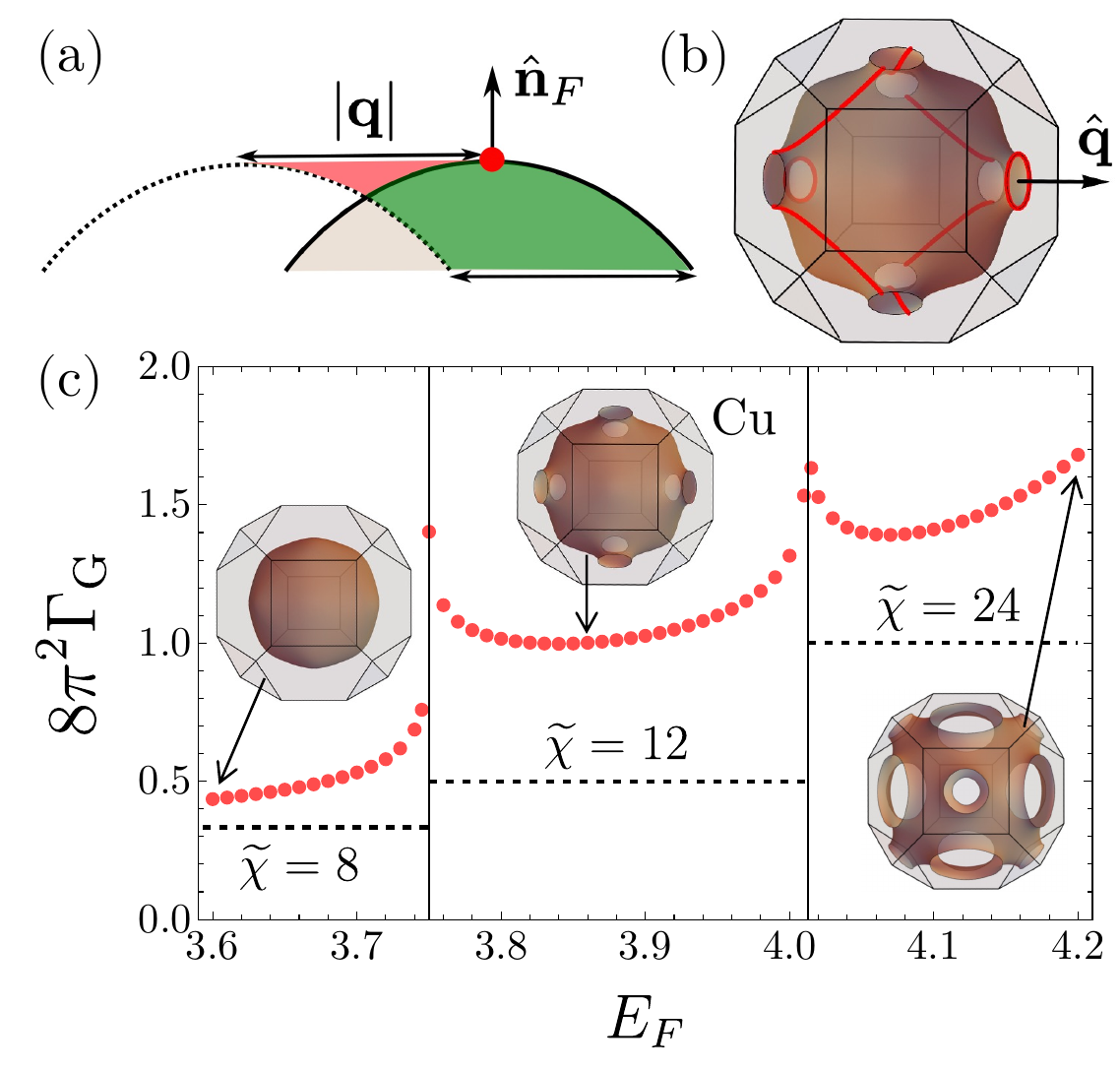}}
    \caption{(a) Structure factor $S(\bq)$ in Eq.~\eqref{eq: single band Sq} measures the volume of the green-shaded region between the Fermi surface (solid) and its shifted counterpart (dotted). $S^{(1)}(\bq)$ corresponds to the total shaded region, which overcounts a sliver of volume (red), $S^{(3)}(\bq)$, around Fermi surface critical points where $\bq\cdot \hat{\bn}_F=0$. (b) Critical lines on the Fermi surface of copper are indicated (red) for a specific orientation of $\bq$. (c) $\Gamma_{{\rm G}}$ as a function of $E_F$ for the model in Eq. \eqref{eq: Cooper FS model}, showing three topological classes of Fermi surface. Dashed lines indicate the corresponding topological lower bounds given by Eq.~\eqref{eq: topological bound on Gamma_G}.}
    \label{fig: geometric contribution}
\end{figure}

\noindent{\color{blue}\emph{Geometry of Fermi surfaces--}} 
The geometric interpretation of Eq.~\eqref{eq: single band Sq} implies that the relevant volume difference is proportional to $\abs{\bq}$ at the leading order, with a coefficient given by the area of the Fermi surface projected onto the plane perpendicular to $\hat{\bq}$. 
Following Eq.~\eqref{eq: expansion of Sq} \cite{supp}, 
\begin{equation}\label{eq: alpha}
    S^{(1)}(\bq) = \frac{1}{16\pi^3} \sum_m \int_{{\rm FS}_m} dA_F\;\abs{ \bq\cdot \hat{\bn}_F},
\end{equation}
where $dA_F>0$ is the area element of the Fermi surface, $\hat{\bn}_F$ is the unit outward normal, and the sum runs over all disconnected sheets of the Fermi surface indexed by $m$. 

Importantly, as illustrated in Fig.~\ref{fig: geometric contribution}(a),  $S^{(1)}(\bq)$ has \textit{overcounted} (hence the minus sign in Eq.~\eqref{eq: expansion of Sq}) a thin sliver of volume of order $\abs{\bq}^3$ surrounding every \textit{critical line} on the Fermi surface where $\bq\cdot \hat{\bn}_F=0$. Specifying an arbitrary $\hat{k}_z$ direction perpendicular to $\bq$ and considering the cross-section of the Fermi sea on a fixed $k_z$ plane, the Fermi surface around a critical point $p$ can be parametrized as $k_y=k_x^2/[2r_p(k_z)]$, where $r_p(k_z)$ is the local radius of curvature along that plane with $k_x$ parallel to $\bq$. Correspondingly, the over-counted area on that plane is $\abs{\bq}^3/(24\abs{r_p(k_z)})$. Integrating along the $k_z$-direction, we obtain
\begin{equation}\label{eq: gamma_G 1}
    S^{(3)}_{\rm G}(\bq) = \frac{\abs{\bq}^3}{192\pi^3}\int dk_z \sum_{p\in \text{crit.}} \frac{1}{\abs{r_p(k_z)}}.
\end{equation}
Alternatively, as detailed in the supplemental material (SM) \cite{supp}, $S^{(3)}_{\rm G}$ can be recast as a Fermi surface integral,
\begin{subequations}\label{eq: gamma_G 2}
\begin{align}
    S^{(3)}_{\rm G}(\bq) &= \frac{1}{192\pi^3} \sum_m \int_{{\rm FS}_m} dA_F\delta(\bq\cdot\hat{\bn}_F)I(\bs, \bq),\\
    I(\bs,\bq) &= [K^{\mu\nu} (\bq\cdot \partial_\mu \bk_F)(\bq\cdot \partial_\nu \bk_F)]^2,
\end{align}
\end{subequations}
where $\delta(x)$ is the Dirac delta function. Here each sheet of Fermi surface is parametrized as $\bk_F(\bs)$, with a curvilinear coordinate system $\bs \equiv(s^1,s^2)$, $\partial_\mu\equiv \partial/\partial s^\mu$; $g_{\mu\nu}=\partial_\mu\bk_F\cdot\partial_\nu\bk_F$ is the 2D metric induced from the 3D Euclidean $\bk$-space. $K^{\mu\nu} = g^{\mu\sigma}g^{\nu\tau} K_{\sigma\tau}$, where $K_{\mu\nu}(\bs)= \hat{\bn}_F\cdot\partial_\mu\partial_\nu \bk_F$ is the extrinsic curvature tensor on the Fermi surface.

For charge fluctuations with a spherical partition, combining Eqs.~\eqref{eq: def of Gamma} and~\eqref{eq: gamma_G 2}, we need to perform the integral $\int d^2\hat{\bq}\; \delta(\hat{\bq}\cdot\hat{\bn}_F)I(\bs,\hat{\bq})$, which reduces to an angular integral on the plane orthogonal to $\hat{\bn}_F$: $\int_0^{2\pi}d\theta\;\prod_{i=1}^4(\hat{\bq}(\theta)\cdot\partial_{\mu_i}\bk_F) = \frac{\pi}{4}(g_{\mu_1\mu_2}g_{\mu_3\mu_4}+g_{\mu_1\mu_3}g_{\mu_2\mu_4}+g_{\mu_1\mu_4}g_{\mu_2\mu_3})$. Relating the curvature tensor to the principal curvatures $\kappa_{1,2}$ using $K^{\mu\nu}g_{\mu\nu} = \kappa_1+\kappa_2$ and $K^{\mu\nu}K^{\sigma\tau}g_{\nu\sigma}g_{\tau\mu}=\kappa_1^2+\kappa_2^2$ \cite{supp,frankel2004geometry}, we obtain
\begin{equation}\label{eq: Gamma_G}
    \Gamma_{\rm G} = \frac{1}{48\pi^2} \sum_m\left(\chi_m+\frac{3}{4\pi}\mathcal{W}_m\right).
\end{equation}
The first term with $\chi=\frac{1}{2\pi}\int dA_F\;\kappa_1\kappa_2$ is the integer-valued topological Euler characteristic of the Fermi surface, which can be expressed as $\chi=2-2g$ for a genus-$g$ sheet. The second term with $\mathcal{W} = \frac{1}{4}\int dA_F\; (\kappa_1-\kappa_2)^2$, known as the Willmore energy in differential geometry, measures the anisotropy of Fermi surface. Interestingly, both $\chi$ and $\mathcal{W}$ are invariant under conformal transformations \cite{white1973global}. For a spherical Fermi surface, $\chi=2$ and $\mathcal{W}=0$, Eq. \eqref{eq: gamma and Gamma for spheres} is recovered. An alternative derivation of Eq.~\eqref{eq: Gamma_G} based on Eq.~\eqref{eq: gamma_G 1} is provided in the SM \cite{supp}. Remarkably, as $\kappa_1^2+\kappa_2^2 \geq 2\abs{\kappa_1\kappa_2}$, Eq.~\eqref{eq: Gamma_G} implies a topological bound
\begin{equation}\label{eq: topological bound on Gamma_G}
    \Gamma_{\rm G} \geq \frac{1}{192\pi^2}\widetilde{\chi}\;, \quad \widetilde{\chi}\equiv \sum_m(\chi_m+3\abs{\chi_m}),
\end{equation}
where $\widetilde{\chi}\in\mathbb {Z}^+_0$ is a non-negative topological invariant. This inequality is saturated only when every sheet of the Fermi surface is isotropic. It is reasonable to expect that the inequality can be strengthened further. For example, it is known that for a torus ($\chi=0$) immersed in $\mathbb{R}^3$, $\mathcal{W}\ge2\pi^2$ \cite{marques2014min}.

Equations~\eqref{eq: gamma_G 1},~\eqref{eq: gamma_G 2},~\eqref{eq: Gamma_G}, together with the above Euler characteristic bound constitute the main results for the first part of our work on the geometry of Fermi surface. Notably, while the topology of Fermi surface has been previously connected to multi-point density correlations \cite{TamKane2023b} and multipartite fluctuations \cite{TamKane2022a}, here we have identified a refined geometrical characterization of Fermi surfaces in terms of the structure factor and bipartite charge fluctuations. Next we provide an example associated with realistic materials to illustrate the content of this newly identified geometric quantity.

\noindent{\color{blue}\emph{Example 1--}}
The shape of the Fermi surface in copper is captured by \cite{roaf1962fermi, copper},
\begin{equation}\label{eq: Cooper FS model}
\begin{split}
    &E_\bk = (3-\cos k_x\cos k_y-...)+C_{200}(3-\\
    &\cos2 k_x-...)+C_{220}(3-\cos2 k_x\cos2 k_y-...)+\\
    &C_{310}(6-\cos3 k_x\cos k_y-\cos k_x\cos3 k_y-...),
\end{split}
\end{equation}
where ellipsis indicate terms implied by the cubic symmetry,  $C_{200}=0.12$, $C_{220}=0.055$, and $C_{310}=0.0015$. The Fermi surface of copper matching the de Haas-van Alphen oscillation measurements is given by $E_\bk=E_F = 3.86$, as depicted in Fig. \ref{fig: geometric contribution}(b), which has genus $g=4$ ($\chi=-6$), and is topologically equivalent to the Fermi surfaces in silver and gold \cite{roaf1962fermi}. Figure \ref{fig: geometric contribution}(c) shows the behavior of $\Gamma_{\rm G}$ as $E_F$ is varied to cover three topological classes ($\chi=2$, $-6$ and $-12$). As expected, $\Gamma_{\rm G}$ approaches the topological bound only when the Fermi surface becomes isotropic with $\chi=2$ (the Fermi surface of sodium lies in this class \cite{lee1966haas}). On the other hand, $\Gamma_{\rm G}$ peaks at Lifshitz transitions where the topology changes such that parts of the Fermi surface become singular.

\noindent{\color{blue}\emph{Quantum geometry of Fermi surfaces--}} We now incorporate quantum geometric effects associated with the geometry of wavefunctions on the Fermi surface. Following Eq.~\eqref{eq: multiband Sq} and expanding $\mathscr{P}_{m,n}(\bk,\bq)$ in powers of $q$, we first notice that the order-$q$ term vanishes: $q_a\Tr[P_m\partial^aP_n]=0$ \cite{order_q}, where $\partial^a\equiv \partial/\partial k_a$. The leading quantum geometric effect thus arises at order $q^2$, resulting in the $S^{(2)}(\bq)$ term in Eq.~\eqref{eq: expansion of Sq}. This is nonetheless \textit{not} a property of the Fermi \textit{surface} but of the entire filled Fermi sea, which can be non-zero even in an insulator with $S^{(2)}(\bq) = q_a q_b \int_{\bk \in \text{BZ}}\mathcal{G}^{ab}$, where $\mathcal{G}^{ab}\equiv \frac{1}{2}\Tr[\partial^aP\partial^bP]$ is the quantum metric of all occupied bands ($P = \sum_{m\in\text{occ.}}P_m$) \cite{Regnault2013, Onishi2025, Tam_corner2024, Wu_corner2025, Abbamonte2025}. 

To extract Fermi surface properties, we again resort to the singular $\abs{\bq}^3$ term. Using the identity $f_{m,\bk}(1-f_{n,\bk+\bq}) = (f_{m,\bk}-f_{n,\bk+\bq})\theta(E_{n,\bk+\bq}-E_{m,\bk})$, and under the general assumption that Fermi surfaces from distinct bands (and distinct components) do not intersect, it can be seen that non-analyticity in $\bq$ only arises when $m=n$. Expressing $f_{m,\bk}-f_{m,\bk+\bq} = \bq\cdot\hat{\bn}_Fv_F\delta(E_F-E_{m,\bk})$, with $v_F$ the magnitude of Fermi velocity, we can combine this with the $q^2$-term in $F_{m,m}(\bk,\bq)$ and obtain
\begin{equation}
    S^{(3)}_{\rm QG}(\bq) = \frac{1}{16\pi^3}\sum_m \int_{{\rm FS}_m} dA_F\; \abs{\bq\cdot\hat{\bn}_{F}}\;
    \mathcal{G}_m^{ab}q_a q_b
    \label{eq: gamma QG as FS integral},
\end{equation}
where $\mathcal{G}_m^{ab}(\bk) \equiv \frac{1}{2}\Tr[\partial^aP_m(\bk)\partial^bP_m(\bk)]$ is the quantum metric for the $m$-th band \cite{provost1980riemannian, MarzariVanderbilt1997, Roy2014}.

\begin{figure}[t!]
    \centering
    \resizebox{\columnwidth}{!}{\includegraphics[]{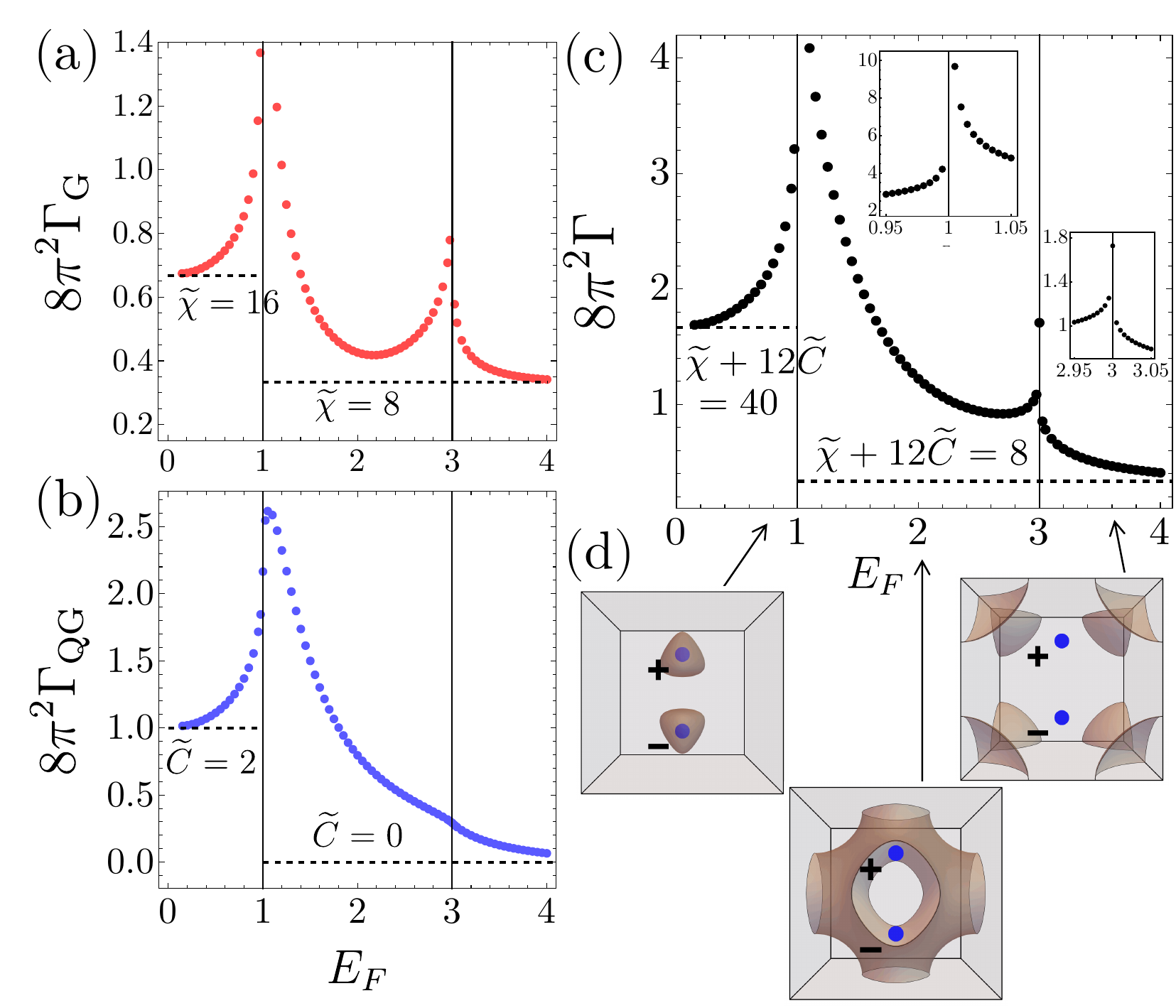}}
    \caption{Interplay between geometry and quantum geometry in a Weyl metal, see Eq.~\eqref{eq: Weyl model}, with $\Gamma_{{\rm G}}$ in (a) and $\Gamma_{{\rm QG}}$ in (b), and $\Gamma=\Gamma_{\rm G}+\Gamma_{\rm QG}$ in (c). Dashed lines indicate the topological lower bounds. (d) Illustration of Fermi surfaces, with the blue dots representing Weyl points with $\pm$-chirality.}
    \label{fig: Weyl}
\end{figure}

The logarithmic coefficient $\Gamma_{\rm QG}$ is then obtained from the angular integral of $S^{(3)}_{\rm QG}(\hat{\bq})$ through the identity $\int d^2\hat{\bq}\;\abs{\hat{\bq}\cdot\hat{\bn}_F}\hat{q}_{a}\hat{q}_{b} = \frac{\pi}{2}(\delta_{ab} +\delta_{ac}\delta_{bd} \hat{n}_F^c \hat{n}_F^d)$, where $\delta_{ab}$ is the flat metric of the 3D Euclidean space.
% and the placement of covariant ($\delta_{ab}$) and contravariant ($\hat{n}_{F}^a$) indices have been kept tracked of. 
Thus
\begin{subequations}\label{eq: Gamma_Q}
\begin{align}
    \Gamma_{{\rm QG}}& = \frac{1}{32\pi^3}\sum_m \int_{{\rm FS}_m} dA_F \left[\delta_{ab}+\hat{n}_{F,a}\hat{n}_{F,b}\right]\mathcal{G}_m^{ab}\\
    = &\frac{1}{32\pi^3}\sum_m \int_{{\rm FS}_m} dA_F \left[\mathcal{G}_m^{\parallel_1 \parallel_1}+\mathcal{G}_m^{\parallel_2 \parallel_2}+2\mathcal{G}_m^{\perp\perp}\right],
\end{align}
\end{subequations}
with $\hat{n}_{F,a}\equiv\delta_{ac}\hat{n}^c_F$.
In the last expression, we have introduced a local orthogonal frame $\{\hat{\bee}^{\parallel_1},\hat{\bee}^{\parallel_2},\hat{\bee}^{\perp}\}$ at every point on the Fermi surface. Finally, recall $\delta_{ab}\mathcal{G}^{ab}_m \geq \abs{\mathcal{F}_m}$ \cite{inequality}, which is sometimes known as the trace inequality \cite{Roy2014}, with $\mathcal{F}_m \equiv \epsilon_{abc} \hat{n}^a_F \mathcal{F}^{bc}_m$ the pseudoscalar Berry curvature on the $m$-th Fermi surface. Introducing the Chern number, $C_m \equiv \frac{1}{2\pi} \int_{\text{FS}_m} dA_F \mathcal{F}_m$, we obtain the following topological bound
\begin{equation}\label{eq: Gamma_Q bound}
    \Gamma_{\rm QG} \geq \frac{1}{16\pi^2}\widetilde{C}, \quad \widetilde{C}\equiv \sum_m \abs{C_m}.
\end{equation}
Equations \eqref{eq: gamma QG as FS integral} and \eqref{eq: Gamma_Q}, together with the above Chern number bound, constitute the main results for the second part regarding the quantum geometric characterization of Fermi surfaces. Combined with Eq. \eqref{eq: topological bound on Gamma_G}, the overall logarithmic coefficient of charge fluctuation satisfies
\begin{equation}\label{eq: total topo bound}
    \Gamma= \Gamma_{\rm G}+\Gamma_{\rm QG} \geq \frac{1}{192\pi^2} \left(\widetilde{\chi}+12\widetilde{C}\right).
\end{equation}
This bound is determined solely by the topology and the quantum topology of Fermi surface. 

\noindent{\color{blue}\emph{Example 2--}} Consider a two-band Hamiltonian exhibiting an isotropic Weyl-point \cite{burkov2018weyl, WeylRMP2018}, $H(\bk) = \bk \cdot \boldsymbol{\sigma}$, 
with the upper ($+$) and lower ($-$) band projectors $P_{\pm}(\bk) = \frac{1}{2}(\mathds{1}\pm \hat{\bk}\cdot\boldsymbol{\sigma})$. The quantum metric is $\mathcal{G}_{\pm}^{ab} =(\delta^{ab}-\hat{k}^a \hat{k}^b)/(4\abs{\bk}^2)$ and the Berry curvature on the isotropic Fermi surface is $\mathcal{F}_{\pm} = \pm 1/(2\abs{\bk}^2)$, which saturates the trace inequality. For a Fermi surface doped away from the Weyl point, $ S^{(3)}_{\rm QG}(\hat{\bq}) = 1/(64\pi^2)$ and  $\Gamma_{\rm QG} = 1/(16\pi^2)$. Together with the geometric contribution (Eq. \eqref{eq: gamma and Gamma for spheres}), we obtain $S^{(3)}(\hat{\bq})=5/(192\pi^2)$ and $\Gamma=5/(48\pi^2)$. It is interesting to contrast this with a semimetal where the Fermi surface diminishes into a Weyl point and is described by a 3+1D CFT, with $S^{(3)}_{\rm CFT}(\hat{\bq}) = 1/(24\pi^2)$ and $\Gamma_{\rm CFT} = 1/(6\pi^2)$ \cite{Mora2019, Wu_fluc_FS}. This discrepancy reflects the non-commutation of two limits, $\abs{\bq}\rightarrow 0 $ and $k_F \rightarrow 0$, and can be understood from the Friedel oscillation in the presence of a finite Fermi surface \cite{supp}.

In Fig.~\ref{fig: Weyl}, we illustrate the interplay between geometry and quantum geometry for a lattice model of a Weyl metal \cite{WeylMinimal2017},
\begin{equation}\label{eq: Weyl model}
    H(\bk) = \sum_{i=x,y}\sigma_i\sin k_i  +\sigma_z\left(2-\sum_{i=x,y,z}\cos k_i\right).
\end{equation}
Varying the Fermi energy $E_F$, the Fermi surface evolves from two electron pockets (each surrounding a Weyl point) to a single genus-3 surface enclosing two Weyl points with opposite chirality, and finally to a hole pocket. The topological bounds are saturated when the Fermi surface approaches the Weyl points (as explained above), or close to the Brillouin zone corner where the quantum geometric effect vanishes and the Fermi surface becomes isotropic.

\noindent{\color{blue}\emph{Ellipsoid partition and beyond--}} Thus far, we have considered $\Gamma$ for a real-space spherical partition. For an ellipsoidal partition surface $\partial A$ specified by $r^a R_{ab}r^b = L^2$ ($R=R^T>0,\;\det R = 1$), we obtain \cite{supp}
\begin{equation}\label{eq: ellipsoid coefficient}
    \Gamma = \frac{1}{\pi}\int'_{\partial A} dA_R\;\abs{\bn_R}^2S^{(3)}(\hat{\bn}_R),
\end{equation}
where the integral $\int'_{\partial A} dA_R$ is performed over the normalized real-space surface $r^a R_{ab} r^b = 1$, $(\bn_R)_a = R_{ab}r^b$ is the outward normal ($\hat{\bn}_R$ is the unit normal). This resembles the Giove-Klich-Widom formula for the leading logarithmically enhanced area-law scaling \cite{GioveKlich2006}, where the coefficient is $\pi^{-1}\int_{\partial A} dA_R S^{(1)}(\hat{\bn}_R)$. As detailed in the SM \cite{supp}, $\Gamma_{\rm G, QG}$ can be cast into the same form as in Eqs.~\eqref{eq: Gamma_G} and~\eqref{eq: Gamma_Q}, only with a deformed Fermi surface $\bk_F \mapsto R^{-1/2}\bk_F$. Equivalently, $\Gamma_{\rm G, QG}$ retain the form of Eqs.~\eqref{eq: Gamma_G} and~\eqref{eq: Gamma_Q} given that the spatial metric is changed from the standard Euclidean metric ($\delta_{ab}$) to $R_{ab}$, under which the ellipsoid is identified as a sphere. Crucially, the topological bounds are left invariant. 

Beyond quadric partitions, a more complicated interplay between real-space and momentum-space is expected. For strictly convex partition surfaces, given an asymptotic expansion of the spatial form factor $\abs{\mathcal{F}_A(\bq)}^2$, which contains the non-oscillatory term $(2\pi)^3\abs{\bq}^{-6}C_A(\hat{\bq})$, the logarithmic coefficient becomes $\Gamma=\int d^2\hat{\bq}\,C_A(\hat{\bq})S^{(3)}(\hat{\bq})$, where $C_A(\hat{\bq})$ encodes the partition geometry. As such, $\Gamma$ can be bounded by combining $\min_{\hat{\bq}}C_A(\hat{\bq})$ with the Fermi surface topological bound in Eq.~\eqref{eq: total topo bound}.

\begin{figure}[t]
    \centering
    \resizebox{0.9\columnwidth}{!}{\includegraphics[]{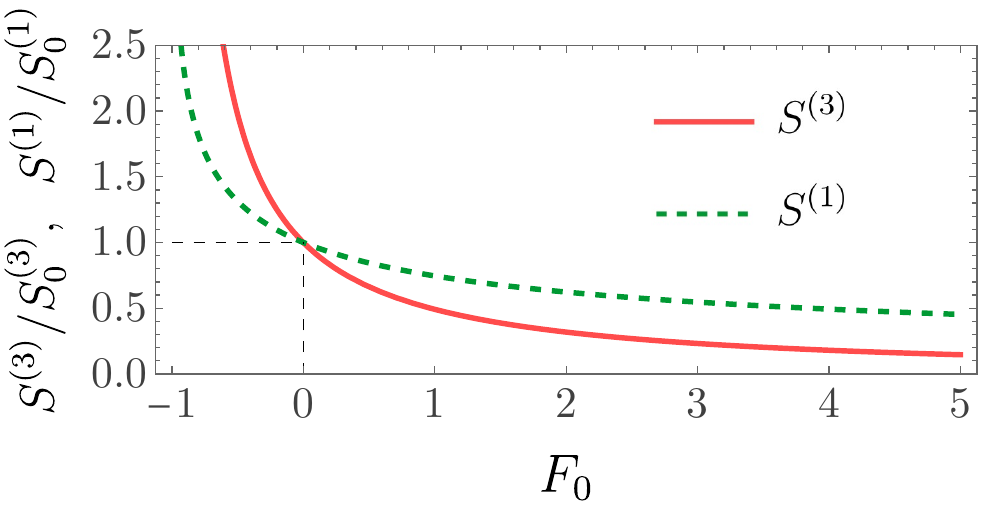}}
    \caption{Coefficients of $S(\bq)$ in isotropic Landau Fermi liquids, where $F_0$ is the $s$-wave channel Landau parameter, $S^{(3)}_0$ and $S^{(1)}_0$ are the coefficients in a free Fermi gas ($F_0=0$).}
    \label{fig: RPA}
\end{figure}
\noindent{\color{blue}\emph{Outlook--}} We have investigated the relationship between charge fluctuations, geometry and quantum geometry of Fermi surfaces in three-dimensional non-interacting metals. An important extension of our work is to incorporate interaction effects. For a one-band isotropic Landau Fermi liquid, using the random phase approximation \cite{supp}, we have shown that $S^{(3)}$ remains in the presence of local interactions, but its value varies upon changing $F_0$, the (dimensionless) Landau parameter in the zeroth angular momentum channel, as shown in Fig.~\ref{fig: RPA}. As $F_0 \rightarrow -1$, $S^{(3)}$ diverges and signals the Pomeranchuk instability for Fermi surface deformation. While there is a history of studying density correlations in interacting fermions \cite{pines1989theory, Giuliani_Vignale_2005, Iwamoto1986, Perdew1991, feldman1998evaluation, metzner1998fermi, Metzner1998PRB,BacheletPRB2000, gori2000correlation, bachelet2001model,Tosi2003, ChubukovMaslov2003, ChubukovMaslov2004,Metzner2006,SwingleSenthil2013,Maslov2018,Torquato2024,Vanoni2025, Wu_fluc_FS, cai2024disorder, daix2025probing, tam2026singular}, our work suggests new ingredients from the geometry and quantum geometry of Fermi surfaces, which may further provide interesting connections to multidimensional bosonization \cite{Haldane1994, Houghton1993, Fradkin1994a, KunYang2012,TanRyu2020, Son2022, cai2024disorder, delacretaz2025symmetry,Wang2025, ye2025berry, Luca2026}. 

Our work also motivates the study of the logarithmic term in entanglement entropy and its connection to quantum anomaly in metals. For non-interacting metals, entanglement entropy is related to the full counting statistics of particle number distribution \cite{KlichLevitov2009, Song2011, Song2012, calabrese2012exact}, and thus the logarithmic term of charge fluctuations contributes to entanglement entropy. One path towards obtaining the logarithmic term in entanglement entropy is by understanding the higher-order charge cumulants, whose connection to Fermi surface geometry remains to be clarified. It is also interesting to ask whether the logarithmic term of the entanglement entropy is stable against interaction effects. Finally, we note that for (3+1)-dimensional CFTs the logarithmic coefficient is related to the geometry of the partition surface as $c_1\chi_{\partial A}+c_2\mathcal{W}_{\partial A}$, where $c_{1,2}$ are related to conformal anomalies \cite{solodukhin2008entanglement, solodukhin2016boundary, nishioka_entanglement_2018}. This resembles the form of $\Gamma_{\rm G}$ in Eq.~\eqref{eq: Gamma_G} and might suggest a duality between the real space partition surface and the momentum space Fermi surface in metallic systems. In this regard, the role of geometry and quantum geometry may provide an interesting perspective for studying Fermi surface anomalies.

\begin{acknowledgments}
\noindent{\color{blue}\emph{Acknowledgments--}} We are grateful to Meng Cheng, Luca Delacrétaz, Jonah Herzog-Arbeitman, Charles Kane, Dmitrii Maslov, Nathan Seiberg, Dam Thanh Son, Carlo Vanoni, Tarik Yefsah and Jiabin Yu for helpful discussions. PMT is supported by a postdoctoral fellowship at the Princeton Center for Theoretical Science and a Croucher Fellowship for Postdoctoral Research. YS is supported by the Adams Fellowship of The Israel Academy of Sciences and Humanities. 
This work is supported by the Gordon and Betty Moore Foundation EPiQS initiative, Grant GBMF8685.01. 
XW and SR are supported by Simons Investigator Grant from the Simons Foundation (Grant No.~566116). FDMH is supported by NSF through the Princeton University (PCCM) Materials Research Science and Engineering Center DMR-2011750.

\end{acknowledgments}

\noindent{\color{blue}\emph{Notes added--}} We would like to draw the reader’s attention to a related paper~\cite{corner_charge_2026}, to appear on the same arXiv listing, which discusses the logarithmic contribution to charge fluctuations in three-dimensional metals for a parallelepiped real-space partition surface, complementary to the smooth quadric partition geometries focused here.

\bibliography{references}

@misc{corner_charge_2026,
  note      = {X.-C. Wu, P. M. Tam, X. Liang, Z. Liu, D.-X. Yao, Z. Yan, and S. Ryu, Corner charge fluctuations in higher dimensions, to appear (2026).}
}

@misc{copper,
note = {We remark that $E_\bk$ in Eq. \eqref{eq: Cooper FS model} does not correspond to the true band structure as multi-band quantum geometric effecs are ignored here. It merely serves as a model for the shape of Fermi surface in copper.}}

@misc{order_q,
note = {This can be understood from $P^2_n=P_n$ and $\Tr[P_n]=1$: $\Tr[P_m \partial^a P_n] = \Tr[P_m \partial^a P^2_n] = 2\delta_{m,n}\Tr[P_n \partial^a P_n]=0$.}}

@misc{inequality,
note = {A simple proof is included for reader's convenience. For any two orthogonal directions ($a,b=1,2$), consider the quantum geometric tensor $\mathcal{Q}^{ab}\equiv \mathcal{G}^{ab} - i\mathcal{F}^{ab} = \Tr[P\partial^a P\partial^b P]$, whose positive semidefiniteness implies $\det(\mathcal{G}) \geq |\mathcal{F}^{12}|^2$. Combined with $(\delta_{ab}\mathcal{G}^{ab})^2 \geq 4\det(\mathcal{G})$, we obtain the stated inequality.}}

@article{daix2025probing,
  title={Probing the Fermi Sea Topology in a Quantum Gas},
  author={Daix, Cyprien and Tam, Pok Man and Dixmerias, Maxime and Verstraten, Joris and de Jongh, Tim and Peaudecerf, Bruno and Kane, Charles L and Yefsah, Tarik},
  journal={arXiv preprint arXiv:2511.23353},
  year={2025},
  url={https://arxiv.org/abs/2511.23353}
}

@article{tam2026singular,
  title={Singular three-point density correlations in two-dimensional Fermi liquids},
  author={Tam, Pok Man and Kane, Charles L},
  journal={arXiv preprint arXiv:2602.16774},
  year={2026},
  url={https://arxiv.org/abs/2602.16774}
}

@book{pines1989theory,
  title={Theory of Quantum Liquids: Normal Fermi Liquids},
  author={Pines, David and Nozi{\`e}res, Philippe},
  year={1989},
  publisher={Advanced Book Classics, Addison-Wesley},
  address={Redwood City, CA},
  series={Advanced Book Classics},
  edition={1st},
  doi={https://doi.org/10.4324/9780429492662}
}

@article{Iwamoto1986,
  title = {Inequalities for frequency-moment sum rules of electron liquids},
  author = {Iwamoto, Naoki},
  journal = {Phys. Rev. A},
  volume = {33},
  issue = {3},
  pages = {1940--1947},
  numpages = {0},
  year = {1986},
  month = {Mar},
  publisher = {American Physical Society},
  doi = {10.1103/PhysRevA.33.1940},
  url = {https://link.aps.org/doi/10.1103/PhysRevA.33.1940}
}

@article{Perdew1991,
  title = {Correlation hole of the spin-polarized electron gas, with exact small-wave-vector and high-density scaling},
  author = {Wang, Yue and Perdew, John P.},
  journal = {Phys. Rev. B},
  volume = {44},
  issue = {24},
  pages = {13298--13307},
  numpages = {0},
  year = {1991},
  month = {Dec},
  publisher = {American Physical Society},
  doi = {10.1103/PhysRevB.44.13298},
  url = {https://link.aps.org/doi/10.1103/PhysRevB.44.13298}
}

@article{metzner1998fermi,
  title={Fermi systems with strong forward scattering},
  author={Metzner, Walter and Castellani, Claudio and Di Castro, Carlo},
  journal={Advances in Physics},
  volume={47},
  number={3},
  pages={317--445},
  year={1998},
  publisher={Taylor \& Francis},
  url={https://doi.org/10.1080/000187398243528}
}

@article{Metzner1998PRB,
  title = {Fermion loops, loop cancellation, and density correlations in two-dimensional Fermi systems},
  author = {Neumayr, Arne and Metzner, Walter},
  journal = {Phys. Rev. B},
  volume = {58},
  issue = {23},
  pages = {15449--15459},
  numpages = {0},
  year = {1998},
  month = {Dec},
  publisher = {American Physical Society},
  doi = {10.1103/PhysRevB.58.15449},
  url = {https://link.aps.org/doi/10.1103/PhysRevB.58.15449}
}

@incollection{feldman1998evaluation,
  title={Evaluation of fermion loops by iterated residues},
  author={Feldman, Joel and Kn{\"o}rrer, Horst and Sinclair, Robert and Trubowitz, Eugene},
  booktitle={Singularities: The Brieskorn Anniversary Volume},
  pages={361--398},
  year={1998},
  publisher={Springer},
  url={https://link.springer.com/content/pdf/10.1007/978-3-0348-8770-0_18.pdf}
}

@article{Metzner2006,
  title = {Fermi surface fluctuations and single electron excitations near Pomeranchuk instability in two dimensions},
  author = {Dell'Anna, Luca and Metzner, Walter},
  journal = {Phys. Rev. B},
  volume = {73},
  issue = {4},
  pages = {045127},
  numpages = {18},
  year = {2006},
  month = {Jan},
  publisher = {American Physical Society},
  doi = {10.1103/PhysRevB.73.045127},
  url = {https://link.aps.org/doi/10.1103/PhysRevB.73.045127}
}

@article{ChubukovMaslov2003,
  title = {Nonanalytic corrections to the Fermi-liquid behavior},
  author = {Chubukov, Andrey V. and Maslov, Dmitrii L.},
  journal = {Phys. Rev. B},
  volume = {68},
  issue = {15},
  pages = {155113},
  numpages = {33},
  year = {2003},
  month = {Oct},
  publisher = {American Physical Society},
  doi = {10.1103/PhysRevB.68.155113},
  url = {https://link.aps.org/doi/10.1103/PhysRevB.68.155113}
}

@article{ChubukovMaslov2004,
  title = {Singular corrections to the Fermi-liquid theory},
  author = {Chubukov, Andrey V. and Maslov, Dmitrii L.},
  journal = {Phys. Rev. B},
  volume = {69},
  issue = {12},
  pages = {121102(R)},
  numpages = {4},
  year = {2004},
  month = {Mar},
  publisher = {American Physical Society},
  doi = {10.1103/PhysRevB.69.121102},
  url = {https://link.aps.org/doi/10.1103/PhysRevB.69.121102}
}

@article{Maslov2018,
  title = {Dynamical susceptibility of a Fermi liquid},
  author = {Zyuzin, Vladimir A. and Sharma, Prachi and Maslov, Dmitrii L.},
  journal = {Phys. Rev. B},
  volume = {98},
  issue = {11},
  pages = {115139},
  numpages = {20},
  year = {2018},
  month = {Sep},
  publisher = {American Physical Society},
  doi = {10.1103/PhysRevB.98.115139},
  url = {https://link.aps.org/doi/10.1103/PhysRevB.98.115139}
}

@article{Tosi2003,
  title = {Analytic theory of ground-state properties of a three-dimensional electron gas with arbitrary spin polarization},
  author = {Davoudi, B. and Asgari, R. and Polini, M. and Tosi, M. P.},
  journal = {Phys. Rev. B},
  volume = {68},
  issue = {15},
  pages = {155112},
  numpages = {9},
  year = {2003},
  month = {Oct},
  publisher = {American Physical Society},
  doi = {10.1103/PhysRevB.68.155112},
  url = {https://link.aps.org/doi/10.1103/PhysRevB.68.155112}
}

@article{BacheletPRB2000,
  title = {Analytic static structure factors and pair-correlation functions for the unpolarized homogeneous electron gas},
  author = {Gori-Giorgi, Paola and Sacchetti, Francesco and Bachelet, Giovanni B.},
  journal = {Phys. Rev. B},
  volume = {61},
  issue = {11},
  pages = {7353--7363},
  numpages = {0},
  year = {2000},
  month = {Mar},
  publisher = {American Physical Society},
  doi = {10.1103/PhysRevB.61.7353},
  url = {https://link.aps.org/doi/10.1103/PhysRevB.61.7353}
}

@article{gori2000correlation,
  title={Correlation energy, pair-distribution functions and static structure factors of jellium},
  author={Gori-Giorgi, Paola and Sacchetti, Francesco and Bachelet, Giovanni B},
  journal={Physica A: Statistical Mechanics and its Applications},
  volume={280},
  number={1-2},
  pages={199--205},
  year={2000},
  publisher={Elsevier}
}

@inproceedings{bachelet2001model,
  title={Model static structure factors and pair-correlation functions for the unpolarized homogeneous electron gas},
  author={Bachelet, Giovanni B and Gori-Giorgi, Paola and Sacchetti, Francesco},
  booktitle={AIP Conference Proceedings},
  volume={577},
  number={1},
  pages={21--37},
  year={2001},
  organization={American Institute of Physics}
}

@article{Torquato2024,
  title = {Correlations in interacting electron liquids: Many-body statistics and hyperuniformity},
  author = {Wang, Haina and Samajdar, Rhine and Torquato, Salvatore},
  journal = {Phys. Rev. B},
  volume = {110},
  issue = {10},
  pages = {104201},
  numpages = {22},
  year = {2024},
  month = {Sep},
  publisher = {American Physical Society},
  doi = {10.1103/PhysRevB.110.104201},
  url = {https://link.aps.org/doi/10.1103/PhysRevB.110.104201}
}

@article{Vanoni2025,
  title = {Quantifying when hyperuniformity of a many-particle system leads to uniformity across length scales},
  author = {Vanoni, Carlo and Steinhardt, Paul J. and Torquato, Salvatore},
  journal = {Phys. Rev. E},
  volume = {112},
  issue = {4},
  pages = {044142},
  numpages = {14},
  year = {2025},
  month = {Oct},
  publisher = {American Physical Society},
  doi = {10.1103/511q-ltf1},
  url = {https://link.aps.org/doi/10.1103/511q-ltf1}
}

@article{burkov2018weyl,
  title={Weyl metals},
  author={Burkov, AA},
  journal={Annual Review of Condensed Matter Physics},
  volume={9},
  pages={359--378},
  year={2018},
  publisher={Annual Reviews},
  doi={https://doi.org/10.1146/annurev-conmatphys-033117-054129}
}

@article{WeylRMP2018,
  title = {Weyl and Dirac semimetals in three-dimensional solids},
  author = {Armitage, N. P. and Mele, E. J. and Vishwanath, Ashvin},
  journal = {Rev. Mod. Phys.},
  volume = {90},
  issue = {1},
  pages = {015001},
  numpages = {57},
  year = {2018},
  month = {Jan},
  publisher = {American Physical Society},
  doi = {10.1103/RevModPhys.90.015001},
  url = {https://link.aps.org/doi/10.1103/RevModPhys.90.015001}
}

@article{WeylMinimal2017,
  title = {Minimal models for topological Weyl semimetals},
  author = {McCormick, Timothy M. and Kimchi, Itamar and Trivedi, Nandini},
  journal = {Phys. Rev. B},
  volume = {95},
  issue = {7},
  pages = {075133},
  numpages = {13},
  year = {2017},
  month = {Feb},
  publisher = {American Physical Society},
  doi = {10.1103/PhysRevB.95.075133},
  url = {https://link.aps.org/doi/10.1103/PhysRevB.95.075133}
}

@book{Giuliani_Vignale_2005, 
place={Cambridge}, 
title={Quantum Theory of the Electron Liquid}, 
publisher={Cambridge University Press}, 
author={Giuliani, Gabriele and Vignale, Giovanni}, 
year={2005},
url={https://doi.org/10.1017/CBO9780511619915}}

@inproceedings{Haldane1994,
  author    = {Haldane, FDM},
  title     = {Luttinger's Theorem and Bosonization of the Fermi Surface},
  booktitle = {Proceedings of the International School of Physics “Enrico Fermi”, Course CXXI:
“Perspectives in Many-Particle Physics”, edited by R. Broglia and J. R. Schrieffer},
  year      = {1994},
  address   = {Amsterdam},
  publisher = {North Holland},
page={5-30},
url={https://arxiv.org/abs/cond-mat/0505529}
}

@article{Houghton1993,
  title = {Bosonization and fermion liquids in dimensions greater than one},
  author = {Houghton, A. and Marston, J. B.},
  journal = {Phys. Rev. B},
  volume = {48},
  issue = {11},
  pages = {7790--7808},
  numpages = {0},
  year = {1993},
  month = {Sep},
  publisher = {American Physical Society},
  doi = {10.1103/PhysRevB.48.7790},
  url = {https://link.aps.org/doi/10.1103/PhysRevB.48.7790}
}

@article{Fradkin1994a,
  title = {Bosonization of the low energy excitations of Fermi liquids},
  author = {Castro Neto, A. H. and Fradkin, Eduardo},
  journal = {Phys. Rev. Lett.},
  volume = {72},
  issue = {10},
  pages = {1393--1397},
  numpages = {0},
  year = {1994},
  month = {Mar},
  publisher = {American Physical Society},
  doi = {10.1103/PhysRevLett.72.1393},
  url = {https://link.aps.org/doi/10.1103/PhysRevLett.72.1393}
}

@article{yu2024non,
  title={Non-trivial quantum geometry and the strength of electron--phonon coupling},
  author={Yu, Jiabin and Ciccarino, Christopher J and Bianco, Raffaello and Errea, Ion and Narang, Prineha and Bernevig, B Andrei},
  journal={Nature Physics},
  volume={20},
  number={8},
  pages={1262--1268},
  year={2024},
  publisher={Nature Publishing Group UK London},
  url={https://doi.org/10.1038/s41567-024-02486-0}
}

@article{berry1984quantal,
  title={Quantal phase factors accompanying adiabatic changes},
  author={Berry, Michael Victor},
  journal={Proceedings of the Royal Society of London. A. Mathematical and Physical Sciences},
  volume={392},
  number={1802},
  pages={45--57},
  year={1984},
  publisher={The Royal Society London},
  url={https://royalsocietypublishing.org/rspa/article/392/1802/45/15579/Quantal-phase-factors-accompanying-adiabatic}
}

@book{shapere1989geometric,
  title={Geometric phases in physics},
  author={Shapere, Alfred and Wilczek, Frank},
  volume={5},
  year={1989},
  publisher={World scientific},
  doi={https://doi.org/10.1142/0613 }
}

@book{bengtsson2017geometry,
  title={Geometry of quantum states: an introduction to quantum entanglement},
  author={Bengtsson, Ingemar and {\.Z}yczkowski, Karol},
  year={2017},
  publisher={Cambridge university press},
  doi={
https://doi.org/10.1017/CBO9780511535048}
}

@book{frankel2004geometry,
  title={The geometry of physics: an introduction},
  author={Frankel, Theodore},
  year={2004},
  publisher={Cambridge university press},
  doi={https://doi.org/10.1017/CBO9781139061377}
}

@article{provost1980riemannian,
  title={Riemannian structure on manifolds of quantum states},
  author={Provost, JP and Vallee, G},
  journal={Communications in Mathematical Physics},
  volume={76},
  pages={289--301},
  year={1980},
  publisher={Springer},
url={https://link.springer.com/article/10.1007/bf02193559}
}

@article{Torma2023essay,
  title = {Essay: Where Can Quantum Geometry Lead Us?},
  author = {T\"orm\"a, P\"aivi},
  journal = {Phys. Rev. Lett.},
  volume = {131},
  issue = {24},
  pages = {240001},
  numpages = {7},
  year = {2023},
  month = {Dec},
  publisher = {American Physical Society},
  doi = {10.1103/PhysRevLett.131.240001},
  url = {https://link.aps.org/doi/10.1103/PhysRevLett.131.240001}
}

@article{yu2025quantum,
  title={Quantum geometry in quantum materials},
  author={Yu, Jiabin and Bernevig, B Andrei and Queiroz, Raquel and Rossi, Enrico and T{\"o}rm{\"a}, P{\"a}ivi and Yang, Bohm-Jung},
  journal={npj Quantum Materials},
  volume={10},
  number={1},
  pages={101},
  year={2025},
  publisher={Nature Publishing Group UK London},
  url={https://www.nature.com/articles/s41535-025-00801-3}
}

@article{HasanKane2010,
  title = {Colloquium: Topological insulators},
  author = {Hasan, M. Z. and Kane, C. L.},
  journal = {Rev. Mod. Phys.},
  volume = {82},
  issue = {4},
  pages = {3045--3067},
  numpages = {0},
  year = {2010},
  month = {Nov},
  publisher = {American Physical Society},
  doi = {10.1103/RevModPhys.82.3045},
  url = {https://link.aps.org/doi/10.1103/RevModPhys.82.3045}
}

@book{moessner2021topological,
  title={Topological phases of matter},
  author={Moessner, Roderich and Moore, Joel E},
  year={2021},
  publisher={Cambridge University Press},
  doi={
https://doi.org/10.1017/9781316226308}
}

@article{Resta2011Rev,
  title={The insulating state of matter: a geometrical theory},
  author={Resta, Raffaele},
  journal={The European Physical Journal B},
  volume={79},
  pages={121--137},
  year={2011},
  publisher={Springer},
  doi={10.1140/epjb/e2010-10874-4}
}

@article{kaganov1979electron,
  title={Electron theory of metals and geometry},
  author={Kaganov, Moisei I and Lifshits, Il'ya Mikhailovich},
  journal={Soviet Physics Uspekhi},
  volume={22},
  number={11},
  pages={904},
  year={1979},
  publisher={IOP Publishing},
  url={https://iopscience.iop.org/article/10.1070/PU1979v022n11ABEH005648/meta}
}

@article{chen2017berry,
  title={Berry Fermi liquid theory},
  author={Chen, Jing-Yuan and Son, Dam Thanh},
  journal={Annals of Physics},
  volume={377},
  pages={345--386},
  year={2017},
  publisher={Elsevier},
  url={https://www.sciencedirect.com/science/article/pii/S0003491616302895}
}

@article{Alexandradinata2018,
  title = {Revealing the Topology of Fermi-Surface Wave Functions from Magnetic Quantum Oscillations},
  author = {Alexandradinata, A. and Wang, Chong and Duan, Wenhui and Glazman, Leonid},
  journal = {Phys. Rev. X},
  volume = {8},
  issue = {1},
  pages = {011027},
  numpages = {18},
  year = {2018},
  month = {Feb},
  publisher = {American Physical Society},
  doi = {10.1103/PhysRevX.8.011027},
  url = {https://link.aps.org/doi/10.1103/PhysRevX.8.011027}
}

@article{alexandradinata2023fermiology,
  title={Fermiology of topological metals},
  author={Alexandradinata, A and Glazman, Leonid},
  journal={Annual Review of Condensed Matter Physics},
  volume={14},
  number={1},
  pages={261--309},
  year={2023},
  publisher={Annual Reviews},
  url={https://www.annualreviews.org/content/journals/10.1146/annurev-conmatphys-040721-021331}
}

@article{NagaosaRMP2010,
  title = {Anomalous Hall effect},
  author = {Nagaosa, Naoto and Sinova, Jairo and Onoda, Shigeki and MacDonald, A. H. and Ong, N. P.},
  journal = {Rev. Mod. Phys.},
  volume = {82},
  issue = {2},
  pages = {1539--1592},
  numpages = {0},
  year = {2010},
  month = {May},
  publisher = {American Physical Society},
  doi = {10.1103/RevModPhys.82.1539},
  url = {https://link.aps.org/doi/10.1103/RevModPhys.82.1539}
}

@article{Dai2020,
  title = {Topological metals induced by the Zeeman effect},
  author = {Sun, Song and Song, Zhida and Weng, Hongming and Dai, Xi},
  journal = {Phys. Rev. B},
  volume = {101},
  issue = {12},
  pages = {125118},
  numpages = {10},
  year = {2020},
  month = {Mar},
  publisher = {American Physical Society},
  doi = {10.1103/PhysRevB.101.125118},
  url = {https://link.aps.org/doi/10.1103/PhysRevB.101.125118}
}

@article{roaf1962fermi,
  title={The Fermi surfaces of copper, silver and gold. II. Calculation of the Fermi surfaces},
  author={Roaf, DJ},
  journal={Philosophical Transactions of the Royal Society of London. Series A, Mathematical and Physical Sciences},
  volume={255},
  number={1052},
  pages={135--152},
  year={1962},
  publisher={The Royal Society London},
url={https://royalsocietypublishing.org/doi/10.1098/rsta.1962.0012}
}

@article{lee1966haas,
  title={The de Haas—van Alphen effect and the Fermi surface of sodium},
  author={Lee, Martin JG},
  journal={Proceedings of the Royal Society of London. Series A. Mathematical and Physical Sciences},
  volume={295},
  number={1443},
  pages={440--457},
  year={1966},
  publisher={The Royal Society London},
  url={https://royalsocietypublishing.org/rspa/article/295/1443/440/12342/The-de-Haas-van-Alphen-effect-and-the-Fermi}
}

@article{Haldane2004,
  title = {Berry Curvature on the Fermi Surface: Anomalous Hall Effect as a Topological Fermi-Liquid Property},
  author = {Haldane, F. D. M.},
  journal = {Phys. Rev. Lett.},
  volume = {93},
  issue = {20},
  pages = {206602},
  numpages = {4},
  year = {2004},
  month = {Nov},
  publisher = {American Physical Society},
  doi = {10.1103/PhysRevLett.93.206602},
  url = {https://link.aps.org/doi/10.1103/PhysRevLett.93.206602}
}

@article{haldane2014attachment,
  title={Attachment of surface" Fermi arcs" to the bulk Fermi surface:" Fermi-level plumbing" in topological metals},
  author={Haldane, FDM},
  journal={arXiv preprint arXiv:1401.0529},
  year={2014},
  url={https://arxiv.org/pdf/1401.0529}
}

@article{Mora2019,
  title = {Bipartite fluctuations and topology of Dirac and Weyl systems},
  author = {Herviou, Lo\"{\i}c and Le Hur, Karyn and Mora, Christophe},
  journal = {Phys. Rev. B},
  volume = {99},
  issue = {7},
  pages = {075133},
  numpages = {19},
  year = {2019},
  month = {Feb},
  publisher = {American Physical Society},
  doi = {10.1103/PhysRevB.99.075133},
  url = {https://link.aps.org/doi/10.1103/PhysRevB.99.075133}
}

@article{cai2024disorder,
  title = {Disorder operators in two-dimensional Fermi and non-Fermi liquids through multidimensional bosonization},
  author = {Cai, Kang-Le and Cheng, Meng},
  journal = {Phys. Rev. B},
  volume = {112},
  issue = {15},
  pages = {155123},
  numpages = {14},
  year = {2025},
  month = {Oct},
  publisher = {American Physical Society},
  doi = {10.1103/d8q8-4r7z},
  url = {https://link.aps.org/doi/10.1103/d8q8-4r7z}
}

@misc{supp,
  note      = {See Supplemental Material for details on: (1) relating $S^{(3)}$ to Fermi surface geometry, (2) relating $S^{(3)}$ to the logarithmic term in charge fluctuations, (3) an alternative derivation for $\Gamma_{\rm{G}}$, (4) structure factor in a Weyl metal, and (5) preliminary analysis on interaction effects in a Fermi liquid.}
}

@article{Regnault2013,
  title = {Geometrical description of fractional Chern insulators based on static structure factor calculations},
  author = {Dobard\ifmmode \check{z}\else \v{z}\fi{}i\ifmmode \acute{c}\else \'{c}\fi{}, E. and Milovanovi\ifmmode \acute{c}\else \'{c}\fi{}, M. V. and Regnault, N.},
  journal = {Phys. Rev. B},
  volume = {88},
  issue = {11},
  pages = {115117},
  numpages = {12},
  year = {2013},
  month = {Sep},
  publisher = {American Physical Society},
  doi = {10.1103/PhysRevB.88.115117},
  url = {https://link.aps.org/doi/10.1103/PhysRevB.88.115117}
}

@article{Onishi2025,
  title = {Quantum weight: A fundamental property of quantum many-body systems},
  author = {Onishi, Yugo and Fu, Liang},
  journal = {Phys. Rev. Res.},
  volume = {7},
  issue = {2},
  pages = {023158},
  numpages = {11},
  year = {2025},
  month = {May},
  publisher = {American Physical Society},
  doi = {10.1103/PhysRevResearch.7.023158},
  url = {https://link.aps.org/doi/10.1103/PhysRevResearch.7.023158}
}

@article{Tam_corner2024,
  title = {Corner Charge Fluctuation as an Observable for Quantum Geometry and Entanglement in Two-Dimensional Insulators},
  author = {Tam, Pok Man and Herzog-Arbeitman, Jonah and Yu, Jiabin},
  journal = {Phys. Rev. Lett.},
  volume = {133},
  issue = {24},
  pages = {246603},
  numpages = {7},
  year = {2024},
  month = {Dec},
  publisher = {American Physical Society},
  doi = {10.1103/PhysRevLett.133.246603},
  url = {https://link.aps.org/doi/10.1103/PhysRevLett.133.246603}
}

@article{Wu_corner2025,
  title = {Corner charge fluctuations and many-body quantum geometry},
  author = {Wu, Xiao-Chuan and Cai, Kang-Le and Cheng, Meng and Kumar, Prashant},
  journal = {Phys. Rev. B},
  volume = {111},
  issue = {11},
  pages = {115124},
  numpages = {25},
  year = {2025},
  month = {Mar},
  publisher = {American Physical Society},
  doi = {10.1103/PhysRevB.111.115124},
  url = {https://link.aps.org/doi/10.1103/PhysRevB.111.115124}
}

@article{Abbamonte2025,
  title = {Quantum entanglement and quantum geometry measured with inelastic x-ray scattering},
  author = {Ba\l{}ut, David and Bradlyn, Barry and Abbamonte, Peter},
  journal = {Phys. Rev. B},
  volume = {111},
  issue = {12},
  pages = {125161},
  numpages = {7},
  year = {2025},
  month = {Mar},
  publisher = {American Physical Society},
  doi = {10.1103/PhysRevB.111.125161},
  url = {https://link.aps.org/doi/10.1103/PhysRevB.111.125161}
}

@article{Wu_fluc_FS,
  title = {Bipartite Fluctuations of Critical Fermi Surfaces},
  author = {Wu, Xiao-Chuan},
  journal = {Phys. Rev. X},
  volume = {15},
  issue = {3},
  pages = {031035},
  numpages = {30},
  year = {2025},
  month = {Aug},
  publisher = {American Physical Society},
  doi = {10.1103/dflw-rksw},
  url = {https://link.aps.org/doi/10.1103/dflw-rksw}
}

@article{SWM2000,
  title = {Polarization and localization in insulators: Generating function approach},
  author = {Souza, Ivo and Wilkens, Tim and Martin, Richard M.},
  journal = {Phys. Rev. B},
  volume = {62},
  issue = {3},
  pages = {1666--1683},
  numpages = {0},
  year = {2000},
  month = {Jul},
  publisher = {American Physical Society},
  doi = {10.1103/PhysRevB.62.1666},
  url = {https://link.aps.org/doi/10.1103/PhysRevB.62.1666}
}

@article{Kivelson1982,
  title = {Wannier functions in one-dimensional disordered systems: Application to fractionally charged solitons},
  author = {Kivelson, S.},
  journal = {Phys. Rev. B},
  volume = {26},
  issue = {8},
  pages = {4269--4277},
  numpages = {0},
  year = {1982},
  month = {Oct},
  publisher = {American Physical Society},
  doi = {10.1103/PhysRevB.26.4269},
  url = {https://link.aps.org/doi/10.1103/PhysRevB.26.4269}
}

@article{MarzariVanderbilt1997,
  title = {Maximally localized generalized Wannier functions for composite energy bands},
  author = {Marzari, Nicola and Vanderbilt, David},
  journal = {Phys. Rev. B},
  volume = {56},
  issue = {20},
  pages = {12847--12865},
  numpages = {0},
  year = {1997},
  month = {Nov},
  publisher = {American Physical Society},
  doi = {10.1103/PhysRevB.56.12847},
  url = {https://link.aps.org/doi/10.1103/PhysRevB.56.12847}
}

@article{Resta1999,
  title = {Electron Localization in the Insulating State},
  author = {Resta, Raffaele and Sorella, Sandro},
  journal = {Phys. Rev. Lett.},
  volume = {82},
  issue = {2},
  pages = {370--373},
  numpages = {0},
  year = {1999},
  month = {Jan},
  publisher = {American Physical Society},
  doi = {10.1103/PhysRevLett.82.370},
  url = {https://link.aps.org/doi/10.1103/PhysRevLett.82.370}
}

@article{Ryu2010MomentumSpaceMetric,
  title = {Momentum space metric, nonlocal operator, and topological insulators},
  author = {Matsuura, Shunji and Ryu, Shinsei},
  journal = {Phys. Rev. B},
  volume = {82},
  issue = {24},
  pages = {245113},
  numpages = {13},
  year = {2010},
  month = {Dec},
  publisher = {American Physical Society},
  doi = {10.1103/PhysRevB.82.245113},
  url = {https://link.aps.org/doi/10.1103/PhysRevB.82.245113}
}

@article{Roy2014,
  title = {Band geometry of fractional topological insulators},
  author = {Roy, Rahul},
  journal = {Phys. Rev. B},
  volume = {90},
  issue = {16},
  pages = {165139},
  numpages = {7},
  year = {2014},
  month = {Oct},
  publisher = {American Physical Society},
  doi = {10.1103/PhysRevB.90.165139},
  url = {https://link.aps.org/doi/10.1103/PhysRevB.90.165139}
}

@article{GioveKlich2006,
  title = {Entanglement Entropy of Fermions in Any Dimension and the Widom Conjecture},
  author = {Gioev, Dimitri and Klich, Israel},
  journal = {Phys. Rev. Lett.},
  volume = {96},
  issue = {10},
  pages = {100503},
  numpages = {4},
  year = {2006},
  month = {Mar},
  publisher = {American Physical Society},
  doi = {10.1103/PhysRevLett.96.100503},
  url = {https://link.aps.org/doi/10.1103/PhysRevLett.96.100503}
}

@article{Swingle2010,
  title = {Entanglement Entropy and the Fermi Surface},
  author = {Swingle, Brian},
  journal = {Phys. Rev. Lett.},
  volume = {105},
  issue = {5},
  pages = {050502},
  numpages = {4},
  year = {2010},
  month = {Jul},
  publisher = {American Physical Society},
  doi = {10.1103/PhysRevLett.105.050502},
  url = {https://link.aps.org/doi/10.1103/PhysRevLett.105.050502}
}

@article{Swingle2012,
  title = {Conformal field theory approach to Fermi liquids and other highly entangled states},
  author = {Swingle, Brian},
  journal = {Phys. Rev. B},
  volume = {86},
  issue = {3},
  pages = {035116},
  numpages = {7},
  year = {2012},
  month = {Jul},
  publisher = {American Physical Society},
  doi = {10.1103/PhysRevB.86.035116},
  url = {https://link.aps.org/doi/10.1103/PhysRevB.86.035116}
}

@article{SwingleSenthil2013,
  title = {Universal crossovers between entanglement entropy and thermal entropy},
  author = {Swingle, Brian and Senthil, T.},
  journal = {Phys. Rev. B},
  volume = {87},
  issue = {4},
  pages = {045123},
  numpages = {13},
  year = {2013},
  month = {Jan},
  publisher = {American Physical Society},
  doi = {10.1103/PhysRevB.87.045123},
  url = {https://link.aps.org/doi/10.1103/PhysRevB.87.045123}
}

@article{KunYang2012,
  title = {Entanglement Entropy of Fermi Liquids via Multidimensional Bosonization},
  author = {Ding, Wenxin and Seidel, Alexander and Yang, Kun},
  journal = {Phys. Rev. X},
  volume = {2},
  issue = {1},
  pages = {011012},
  numpages = {18},
  year = {2012},
  month = {Mar},
  publisher = {American Physical Society},
  doi = {10.1103/PhysRevX.2.011012},
  url = {https://link.aps.org/doi/10.1103/PhysRevX.2.011012}
}

@article{KlichLevitov2009,
  title = {Quantum Noise as an Entanglement Meter},
  author = {Klich, Israel and Levitov, Leonid},
  journal = {Phys. Rev. Lett.},
  volume = {102},
  issue = {10},
  pages = {100502},
  numpages = {4},
  year = {2009},
  month = {Mar},
  publisher = {American Physical Society},
  doi = {10.1103/PhysRevLett.102.100502},
  url = {https://link.aps.org/doi/10.1103/PhysRevLett.102.100502}
}

@article{Song2011,
  title = {Entanglement entropy from charge statistics: Exact relations for noninteracting many-body systems},
  author = {Song, H. Francis and Flindt, Christian and Rachel, Stephan and Klich, Israel and Le Hur, Karyn},
  journal = {Phys. Rev. B},
  volume = {83},
  issue = {16},
  pages = {161408},
  numpages = {4},
  year = {2011},
  month = {Apr},
  publisher = {American Physical Society},
  doi = {10.1103/PhysRevB.83.161408},
  url = {https://link.aps.org/doi/10.1103/PhysRevB.83.161408}
}

@article{Song2012,
  title = {Bipartite fluctuations as a probe of many-body entanglement},
  author = {Song, H. Francis and Rachel, Stephan and Flindt, Christian and Klich, Israel and Laflorencie, Nicolas and Le Hur, Karyn},
  journal = {Phys. Rev. B},
  volume = {85},
  issue = {3},
  pages = {035409},
  numpages = {27},
  year = {2012},
  month = {Jan},
  publisher = {American Physical Society},
  doi = {10.1103/PhysRevB.85.035409},
  url = {https://link.aps.org/doi/10.1103/PhysRevB.85.035409}
}

@article{calabrese2012exact,
  title={Exact relations between particle fluctuations and entanglement in Fermi gases},
  author={Calabrese, Pasquale and Mintchev, Mihail and Vicari, Ettore},
  journal={Europhysics Letters},
  volume={98},
  number={2},
  pages={20003},
  year={2012},
  publisher={IOP Publishing},
doi={10.1209/0295-5075/98/20003},
url={https://iopscience.iop.org/article/10.1209/0295-5075/98/20003/}
}

@article{TanRyu2020,
  title = {Particle number fluctuations, R\'enyi entropy, and symmetry-resolved entanglement entropy in a two-dimensional Fermi gas from multidimensional bosonization},
  author = {Tan, Mao Tian and Ryu, Shinsei},
  journal = {Phys. Rev. B},
  volume = {101},
  issue = {23},
  pages = {235169},
  numpages = {6},
  year = {2020},
  month = {Jun},
  publisher = {American Physical Society},
  doi = {10.1103/PhysRevB.101.235169},
  url = {https://link.aps.org/doi/10.1103/PhysRevB.101.235169}
}

@article{holzhey1994geometric,
  title={Geometric and renormalized entropy in conformal field theory},
  author={Holzhey, Christoph and Larsen, Finn and Wilczek, Frank},
  journal={Nuclear physics b},
  volume={424},
  number={3},
  pages={443--467},
  year={1994},
  publisher={Elsevier},
  doi={10.1016/0550-3213(94)90402-2}
}

@article{calabrese2004entanglement,
  title={Entanglement entropy and quantum field theory},
  author={Calabrese, Pasquale and Cardy, John},
  journal={Journal of statistical mechanics: theory and experiment},
  volume={2004},
  number={06},
  pages={P06002},
  year={2004},
  publisher={IOP Publishing},
  doi={10.1088/1742-5468/2004/06/P06002}
}

@article{FradkinMooire2006,
  title = {Entanglement Entropy of 2D Conformal Quantum Critical Points: Hearing the Shape of a Quantum Drum},
  author = {Fradkin, Eduardo and Moore, Joel E.},
  journal = {Phys. Rev. Lett.},
  volume = {97},
  issue = {5},
  pages = {050404},
  numpages = {4},
  year = {2006},
  month = {Aug},
  publisher = {American Physical Society},
  doi = {10.1103/PhysRevLett.97.050404},
  url = {https://link.aps.org/doi/10.1103/PhysRevLett.97.050404}
}

@article{solodukhin2008entanglement,
  title={Entanglement entropy, conformal invariance and extrinsic geometry},
  author={Solodukhin, Sergey N},
  journal={Physics Letters B},
  volume={665},
  number={4},
  pages={305--309},
  year={2008},
  publisher={Elsevier},
  doi={10.1016/j.physletb.2008.05.071}
}

@article{casini2010entanglement,
  title={Entanglement entropy for the n-sphere},
  author={Casini, H and Huerta, M},
  journal={Physics Letters B},
  volume={694},
  number={2},
  pages={167--171},
  year={2010},
  publisher={Elsevier},
  doi={10.1016/j.physletb.2010.09.054}
}

@article{solodukhin2016boundary,
  title={Boundary terms of conformal anomaly},
  author={Solodukhin, Sergey N},
  journal={Physics Letters B},
  volume={752},
  pages={131--134},
  year={2016},
  publisher={Elsevier},
  url={https://www.sciencedirect.com/science/article/pii/S0370269315008849}
}

@article{Kane2022a,
  title = {{Quantized Nonlinear Conductance in Ballistic Metals}},
  author = {Kane, C. L.},
  journal = {Phys. Rev. Lett.},
  volume = {128},
  issue = {7},
  pages = {076801},
  numpages = {6},
  year = {2022},
  month = {Feb},
  publisher = {American Physical Society},
  doi = {10.1103/PhysRevLett.128.076801},
  url = {https://link.aps.org/doi/10.1103/PhysRevLett.128.076801}
}

@article{TamKane2022a,
  title = {{Topological Multipartite Entanglement in a Fermi Liquid}},
  author = {Tam, Pok Man and Claassen, Martin and Kane, Charles L.},
  journal = {Phys. Rev. X},
  volume = {12},
  issue = {3},
  pages = {031022},
  numpages = {37},
  year = {2022},
  month = {Aug},
  publisher = {American Physical Society},
  doi = {10.1103/PhysRevX.12.031022},
  url = {https://link.aps.org/doi/10.1103/PhysRevX.12.031022}
}

@article{TamKane2022b,
  title = {Probing Fermi Sea Topology by Andreev State Transport},
  author = {Tam, Pok Man and Kane, Charles L.},
  journal = {Phys. Rev. Lett.},
  volume = {130},
  issue = {9},
  pages = {096301},
  numpages = {6},
  year = {2023},
  month = {Mar},
  publisher = {American Physical Society},
  doi = {10.1103/PhysRevLett.130.096301},
  url = {https://link.aps.org/doi/10.1103/PhysRevLett.130.096301}
}

@article{Lifshitz1960,
  title={Anomalies of electron characteristics of a metal in the high pressure region},
  author={Lifshitz, IM and others},
  journal={Sov. Phys. JETP},
  volume={11},
  number={5},
  pages={1130--1135},
  year={1960},
  url={http://www.jetp.ras.ru/cgi-bin/e/index/e/11/5/p1130?a=list}
}

@article{Yang2022,
  doi = {10.22331/q-2022-11-10-857},
  url = {https://doi.org/10.22331/q-2022-11-10-857},
  title = {Quantized {N}onlinear {T}ransport with {U}ltracold {A}toms},
  author = {Yang, Fan and Zhai, Hui},
  journal = {{Quantum}},
  issn = {2521-327X},
  publisher = {{Verein zur F{\"{o}}rderung des Open Access Publizierens in den Quantenwissenschaften}},
  volume = {6},
  pages = {857},
  month = nov,
  year = {2022}
}

@article{Zhang2023,
  title = {Quantized topological response in trapped quantum gases},
  author = {Zhang, Pengfei},
  journal = {Phys. Rev. A},
  volume = {107},
  issue = {3},
  pages = {L031305},
  numpages = {5},
  year = {2023},
  month = {Mar},
  publisher = {American Physical Society},
  doi = {10.1103/PhysRevA.107.L031305},
  url = {https://link.aps.org/doi/10.1103/PhysRevA.107.L031305}
}

@article{Bakr_review,
	author = {Gross, Christian and Bakr, Waseem S.},
	date = {2021/12/01},
	doi = {10.1038/s41567-021-01370-5},
	id = {Gross2021},
	isbn = {1745-2481},
	journal = {Nature Physics},
	number = {12},
	pages = {1316--1323},
	title = {Quantum gas microscopy for single atom and spin detection},
	url = {https://doi.org/10.1038/s41567-021-01370-5},
	volume = {17},
	year = {2021}}

@article{TamKane2023a,
  title = {Topological Andreev rectification},
  author = {Tam, Pok Man and De Beule, Christophe and Kane, Charles L.},
  journal = {Phys. Rev. B},
  volume = {107},
  issue = {24},
  pages = {245422},
  numpages = {19},
  year = {2023},
  month = {Jun},
  publisher = {American Physical Society},
  doi = {10.1103/PhysRevB.107.245422},
  url = {https://link.aps.org/doi/10.1103/PhysRevB.107.245422}
}

@book{Giamarchi2004,
  title={Quantum Physics in One Dimension},
  author={Giamarchi, Thierry},
  isbn={9780198525004},
  lccn={2004299020},
  series={International Series of Monographs on Physics},
  url={https://books.google.com/books?id=1MwTDAAAQBAJ},
  year={2004},
  publisher={Clarendon Press}
}

@article{Son2022,
  title = {Nonlinear bosonization of Fermi surfaces: The method of coadjoint orbits},
  author = {Delacr\'etaz, Luca V. and Du, Yi-Hsien and Mehta, Umang and Son, Dam Thanh},
  journal = {Phys. Rev. Res.},
  volume = {4},
  issue = {3},
  pages = {033131},
  numpages = {26},
  year = {2022},
  month = {Aug},
  publisher = {American Physical Society},
  doi = {10.1103/PhysRevResearch.4.033131},
  url = {https://link.aps.org/doi/10.1103/PhysRevResearch.4.033131}
}

@article{delacretaz2025symmetry,
  title={Symmetry and causality constraints on Fermi liquids},
  author={Delacr{\'e}taz, Luca V and Chowdhury, Subham Dutta and Mehta, Umang},
  journal={Journal of High Energy Physics},
  volume={2025},
  number={10},
  pages={1--60},
  year={2025},
  publisher={Springer},
  url={https://link.springer.com/article/10.1007/JHEP10(2025)171}
}

@article{Luca2026,
  title = {Quantizing bosonized Fermi surfaces},
  author = {Chen, Sihan and Delacr\'etaz, Luca V.},
  journal = {Phys. Rev. B},
  volume = {113},
  issue = {12},
  pages = {125150},
  numpages = {23},
  year = {2026},
  month = {Mar},
  publisher = {American Physical Society},
  doi = {10.1103/zdtz-mdtn},
  url = {https://link.aps.org/doi/10.1103/zdtz-mdtn}
}

@article{Wang2025,
  title = {Coadjoint-orbit bosonization of a Fermi surface in a weak magnetic field},
  author = {Ye, Mengxing and Wang, Yuxuan},
  journal = {Phys. Rev. B},
  volume = {112},
  issue = {3},
  pages = {035113},
  numpages = {14},
  year = {2025},
  month = {Jul},
  publisher = {American Physical Society},
  doi = {10.1103/4yfv-jct8},
  url = {https://link.aps.org/doi/10.1103/4yfv-jct8}
}

@article{ye2025berry,
  title={Berry curvature and quantum oscillation from multi-orbital coadjoint-orbit bosonization},
  author={Ye, Mengxing and Wang, Yuxuan},
  journal={SciPost Physics},
  volume={19},
  number={3},
  pages={078},
  year={2025},
  url={https://www.scipost.org/SciPostPhys.19.3.078}
}

@article{TamKane2023b,
  title = {Topological density correlations in a Fermi gas},
  author = {Tam, Pok Man and Kane, Charles L.},
  journal = {Phys. Rev. B},
  volume = {109},
  issue = {3},
  pages = {035413},
  numpages = {14},
  year = {2024},
  month = {Jan},
  publisher = {American Physical Society},
  doi = {10.1103/PhysRevB.109.035413},
  url = {https://link.aps.org/doi/10.1103/PhysRevB.109.035413}
}

@article{Yang2023,
  title = {Euler-Chern correspondence via topological superconductivity},
  author = {Yang, Fan and Li, Xingyu and Li, Chengshu},
  journal = {Phys. Rev. Res.},
  volume = {5},
  issue = {3},
  pages = {033073},
  numpages = {6},
  year = {2023},
  month = {Aug},
  publisher = {American Physical Society},
  doi = {10.1103/PhysRevResearch.5.033073},
  url = {https://link.aps.org/doi/10.1103/PhysRevResearch.5.033073}
}

@article{Jia2025,
  title = {Generic reduction theory for Fermi sea topology in metallic systems},
  author = {Jia, Wei},
  journal = {Phys. Rev. B},
  volume = {111},
  issue = {15},
  pages = {155115},
  numpages = {11},
  year = {2025},
  month = {Apr},
  publisher = {American Physical Society},
  doi = {10.1103/PhysRevB.111.155115},
  url = {https://link.aps.org/doi/10.1103/PhysRevB.111.155115}
}

@article{white1973global,
  title={A global invariant of conformal mappings in space},
  author={White, James H},
  journal={Proceedings of the American Mathematical Society},
  volume={38},
  number={1},
  pages={162--164},
  year={1973},
  url={https://www.jstor.org/stable/2038790}
}

@article{marques2014min,
  title={Min-max theory and the Willmore conjecture},
  author={Marques, Fernando C and Neves, Andr{\'e}},
  journal={Annals of mathematics},
  pages={683--782},
  year={2014},
  publisher={JSTOR},
  url={https://annals.math.princeton.edu/2014/179-2/p06},
  doi={https://doi.org/10.4007/annals.2014.179.2.6}
}

@article{nishioka_entanglement_2018,
  title = {Entanglement entropy: Holography and renormalization group},
  author = {Nishioka, Tatsuma},
  journal = {Rev. Mod. Phys.},
  volume = {90},
  issue = {3},
  pages = {035007},
  numpages = {56},
  year = {2018},
  month = {Sep},
  publisher = {American Physical Society},
  doi = {10.1103/RevModPhys.90.035007},
  url = {https://link.aps.org/doi/10.1103/RevModPhys.90.035007}
}

@book{schulke2007electron,
  title={Electron dynamics by inelastic X-ray scattering},
  author={Sch{\"u}lke, Winfried},
  volume={7},
  year={2007},
  publisher={OUP Oxford},
  url={https://global.oup.com/academic/product/electron-dynamics-by-inelastic-x-ray-scattering-9780198510178?cc=us&lang=en&}
}

@article{nelson2007imaging,
  title={Imaging single atoms in a three-dimensional array},
  author={Nelson, Karl D and Li, Xiao and Weiss, David S},
  journal={Nature Physics},
  volume={3},
  number={8},
  pages={556--560},
  year={2007},
  publisher={Nature Publishing Group UK London},
  url={https://www.nature.com/articles/nphys645}
}

@article{3DQGM2020,
  title = {Spatial tomography of individual atoms in a quantum gas microscope},
  author = {El\'{\i}asson, Ott\'o and Laustsen, Jens S. and Heck, Robert and M\"uller, Romain and Arlt, Jan J. and Weidner, Carrie A. and Sherson, Jacob F.},
  journal = {Phys. Rev. A},
  volume = {102},
  issue = {5},
  pages = {053311},
  numpages = {7},
  year = {2020},
  month = {Nov},
  publisher = {American Physical Society},
  doi = {10.1103/PhysRevA.102.053311},
  url = {https://link.aps.org/doi/10.1103/PhysRevA.102.053311}
}

\clearpage
\newpage

\clearpage
\newpage
\widetext
\begin{center}
\textbf{\large Supplemental Materials for ``Fermi Surface Geometry from Charge Fluctuations in Three-Dimensional Metals"}\\
\vspace{0.5cm}
\text{Pok Man Tam, Yarden Sheffer, Xiao-Chuan Wu, F. D. M. Haldane, and Shinsei Ryu}
\end{center}
\onecolumngrid
\setcounter{secnumdepth}{2}
\renewcommand{\theequation}{\thesection.\arabic{equation}}
\renewcommand{\theHequation}{\theHsection.\arabic{equation}}
\renewcommand{\thefigure}{\thesection.\arabic{figure}}  

The supplemental information consists of five sections. In Sec.~\ref{supp_sec: derivation} we provide detailed derivations of $S^{(3)}_{\rm G}(\bq)$ and $S^{(3)}_{\rm QG}(\bq)$, which give Eqs.~\eqref{eq: gamma_G 2} and~\eqref{eq: gamma QG as FS integral} in the main text. In Sec.~\ref{supp_sec: bipartite fluctuation} we discuss the bipartite charge fluctuations in connection to the structure factor $S(\bq)$, and derive the logarithmic coefficient of the finite-size scaling for general ellipsoid partition surfaces, as well as the topological bounds. In Sec.~\ref{supp_sec: Gamma_G} we provide an elementary derivation for $\Gamma_{\rm G}$ from Eq.~\eqref{eq: gamma_G 1}. In Sec.~\ref{supp_sec: Weyl}, we analyze the structure factor in a Weyl metal, showing that the discrepancy between the semimetallic and metallic cases, discussed in the main text (Example 2), can be understood from the effect of Friedel oscillation. In Sec.~\ref{supp_sec: FL}, we discuss interaction effects on $S(\bq)$ for a one-band isotropic Fermi liquid. 

\section{Deriving the $\abs{\bq}^3$-term in $S(\bq)$}\label{supp_sec: derivation}
\setcounter{equation}{0}
\setcounter{figure}{0} 

The central interest of this work is the structure factor and charge fluctuations in metallic systems with Fermi surfaces. The connection between the structure factor and charge fluctuations is explained in Sec. \ref{supp_sec: bipartite fluctuation}. Here we focus on the detailed aspect of the structure factor in connection to the geometry and quantum geometry of Fermi surfaces.\\

To set the stage, we consider a lattice system consists of point-like fermionic orbitals ($\ket{\bR,\sigma} = c^\dagger_{\bR,\sigma}\ket{0}$, $\bR$ labels the unit cell, $\sigma$ labels the intra-cell orbital, and $\bR_\sigma$ is the physical position of the orbital), governed by a Hamiltonian with a lattice translational symmetry:
\begin{equation}
    \mathcal{H} = \sum_{\bR, \bR'}\sum_{\sigma,\sigma'} h_{\sigma,\sigma'}(\bR-\bR') c^\dagger_{\bR,\sigma} c_{\bR',\sigma'}.
\end{equation}
We assume periodic boundary conditions and denote the number of unit cells by $N_c$. 
The Bloch Hamiltonian at momentum $\bk$, $H_{\sigma,\sigma'}(\bk) = N_c^{-1}\sum_{\bR,\bR'} e^{-i\bk\cdot\bR_\sigma}h_{\sigma,\sigma'}(\bR-\bR')e^{i\bk\cdot\bR'_{\sigma'}}$, is diagonalized by the Bloch eigenvectors $U_{\sigma,m}(\bk)$ as $\sum_{\sigma,\sigma'}U^*_{\sigma,m}(\bk)H_{\sigma,\sigma'}(\bk)U_{\sigma',n}(\bk) = \delta_{m,n} E_{m,\bk}$, where $E_{m,\bk}$ is the energy of the $m$-th band. The eigenvectors are normalized as $\sum_\sigma U^*_{\sigma,m}(\bk)U_{\sigma,n}(\bk) = \delta_{m,n}$. Introducing the creation operator for the $m$-th band at momentum $\bk$ inside the Brillouin zone (BZ) as
\begin{equation}\label{supp_eq: m-th band creation operator, lattice}
    c^\dagger_{m,\bk} = \frac{1}{\sqrt{N_c}} \sum_{\bR,\sigma} e^{i\bk\cdot \bR_\sigma}U_{\sigma,m}(\bk)c^\dagger_{\bR,\sigma},
\end{equation}
we have $\mathcal{H} = \sum_{\bk\in \rm BZ} \sum_mE_{m,\bk} c^\dagger_{m,\bk} c_{m,\bk}$. The momentum-space fermion density operator can be expressed as
\begin{equation}\label{supp_eq: general rho expression}
    \rho_\bq \equiv \int d^3\br e^{-i\bq\cdot\br}\rho(\br) = \sum_{\bR,\sigma} e^{-i\bq\cdot\bR_\sigma} c^\dagger_{\bR,\sigma} c_{\bR,\sigma}  = \sum_{\bk\in\rm BZ}\sum_{\sigma, m, n} U^*_{\sigma,m}(\bk)U_{\sigma, n}(\bk+\bq) c^\dagger_{m,\bk} c_{n,\bk+\bq},
\end{equation}
where $\rho(\br) = \sum_{\bR,\sigma}\delta(\br-\bR_\sigma)c^\dagger_{\bR,\sigma}c_{\bR,\sigma}$ is the density in real space. This gives Eq.~\eqref{eq: eq: general form for density} in the main text. Correspondingly, the structure factor can be expressed as
\begin{equation}\label{supp_eq: unsymmetrized Sq}
    S(\bq) \equiv \frac{\cc{\rho_\bq \rho_{-\bq}}}{V} = \int_{\rm BZ} \frac{d^3\bk}{(2\pi)^3}\sum_{m,n} \mathscr{P}_{m,n}(\bk, \bq) f_{m,\bk}(1-f_{n,\bk+\bq}),\quad \mathscr{P}_{m,n}(\bk, \bq)\equiv\Tr[P_{m}(\bk) P_{n}(\bk+\bq)], 
\end{equation}
where $V$ is the total system volume and the thermodynamic limit is taken in the final result. In the above, we have defined $[P_m(\bk)]_{\sigma,\sigma'} \equiv U_{\sigma, m}(\bk)U^*_{\sigma',m}(\bk)$ as the $m$-th band projector, $\Tr$ as the trace over the orbital space, and $f_{m,\bk} = \theta(E_F-E_{m,\bk})$ is the zero-temperature Fermi occupation for the $m$-th band (with $E_F$ the Fermi energy). This gives Eq. \eqref{eq: multiband Sq} in the main text. Notice that $S(\bq) = S(-\bq)$, and using the identity $f_{m,\bk}(1-f_{n,\bk+\bq}) = (f_{m,\bk}-f_{n,\bk+\bq})\theta(E_{n,\bk+\bq}-E_{m,\bk})$, we could as well express 
\begin{equation}\label{supp_eq: symmetrized Sq}
\begin{split}
    S(\bq) = \frac{1}{2}\int_{\rm BZ} \frac{d^3\bk}{(2\pi)^3} \sum_{m,n} \mathscr{P}_{m,n}(\bk, \bq)(f_{m,\bk}-f_{n,\bk+\bq}) \sgn(E_{n,\bk+\bq}-E_{m,\bk}).
\end{split}
\end{equation}
In passing, we note that the formulation in terms of ``point-like" orbitals is adopted only for simplicity. In terms of the continuum Bloch wavefunction $\psi_{m,\bk}(\br) = e^{i\bk\cdot\br} u_{m,\bk}(\br)$ (for the $m$-th band at momentum $\bk$, with $u_{m,\bk}(\br)$ the cell-periodic part of the Bloch function), we can define the band creation operator (in analogue to Eq.~\eqref{supp_eq: m-th band creation operator, lattice}) as $c^\dagger_{m,\bk}=\int d^3\br\psi_{m,\bk}(\br)c^\dagger_{\br}$, where $c^\dagger_\br$ creates a fermion at position $\br$ in the continuous real space. Correspondingly, the density operator can be expressed as $\rho_\bq = \int d^3\br e^{-i\bq\cdot\br}c^\dagger_\br c_\br  =\sum_{m,n}\sum_{\bk \in \rm BZ} \left(\int d^3\br\; u^*_{m,\bk}(\br)u_{n,\bk+\bq}(\br)\right) c^\dagger_{m,\bk} c_{n,\bk+\bq}$, and the structure factor again takes the form of Eqs.~\eqref{supp_eq: unsymmetrized Sq} and ~\eqref{supp_eq: symmetrized Sq}, with the following replacement:
\begin{equation}\label{supp_eq: continuum formulation for projector}
    \mathscr{P}_{m,n}(\bk, \bq)\equiv\Tr[P_{m}(\bk) P_{n}(\bk+\bq)] \rightarrow \abs{\int d^3\br\; u^*_{m,\bk}(\br)u_{n,\bk+\bq}(\br)}^2.
\end{equation}
Our results hold irrespective of which formulation we adopt. \\

In this work, we are interested in the non-analytic $\abs{\bq}$-dependence (particularly the $\abs{\bq}^3$-term):
\begin{equation}\label{supp_eq: expansion of Sq}
    S(\bq) = S^{(1)}(\bq) - S^{(3)}(\bq)+...\;,\quad S^{(d)}(\bq) =S^{(d)}(\hat{\bq})\abs{\bq}^d,
\end{equation}
with $\hat{\bq}\equiv \bq/\abs{\bq}$ and the ellipsis represent higher order or regular parts. The minus sign put before $S^{(3)}$ is intentional, as we will explain in Sec.~\ref{supp_sec: geometric}, such that $S^{(3)}\geq 0$. As we shall see, $S^{(1)}$ is related to the leading logarithmically enhanced area-law divergence in bipartite charge fluctuations, while $S^{(3)}$ is related to the universal subleading logarithmic divergence in charge fluctuations. The central interest here is the relation between $S^{(3)}$ and the geometry (as well as quantum geometry) of the Fermi surface.\\

The leading $\abs{\bq}$-term in Eq. \eqref{supp_eq: symmetrized Sq} is completely geometrical, obtained from the leading term in the expansion of  $\mathscr{P}_{m,n}(\bk, \bq) = \delta_{m,n} +\mathcal{O}(q)$ and $(f_{m,\bk}-f_{m,\bk+\bq}) \sgn(E_{m,\bk+\bq}-E_{m,\bk}) = \abs{\bq\cdot \bv_{m,\bk}}\delta(E_F-E_{m,\bk})$, with $\bv_{m,\bk}=\nabla_\bk E_{m,\bk}$. Thus,
\begin{equation}\label{supp_eq: alpha}
    S^{(1)}(\bq) = \frac{1}{16\pi^3} \sum_m \int_{{\rm FS}_m} dA_F\;\abs{\hat{\bn}_F\cdot \bq},
\end{equation}
which is Eq.~\eqref{eq: alpha} in the main text. The above expression sums over all disconnected components of the Fermi surface (indexed here by $m$), with $dA_F$ the area-element of the Fermi surface, and on each point on the Fermi surface we have introduced a unit outward normal $\hat{\bn}_F$. Given any fixed $\bq$, the leading $\abs{\bq}$-coefficient of the structure factor thus measures the cross-sectional area of the Fermi surface projected onto the plane orthogonal to $\bq$. \\

Now, we focus on the $\abs{\bq}^3$-term in Eq. \eqref{supp_eq: symmetrized Sq}. As we will see, there are two contributions to the $\abs{\bq}^3$-term: (1) by keeping $\mathscr{P}_{m,n}(\bk, \bq) = \delta_{m,n}$ and extracting the $\abs{\bq}^3$-term solely from $f_{m,\bk}(1-f_{m, \bk+\bq})$, which gives the \textbf{geometric} contribution denoted as $S^{(3)}_{\rm G}$; (2) combining the second-order term in $\mathscr{P}_{m,n}(\bk, \bq)$ with the $\abs{\bq}$-term from $f_{m,\bk}(1-f_{m, \bk+\bq})$, which gives a \textbf{quantum geometric} contribution denoted as $S^{(3)}_{\rm QG}$. We analyze them separately below. 

\subsection{Geometric contribution: $S^{(3)}_{\rm G}$}\label{supp_sec: geometric}

\subsubsection{Setup}
\begin{figure}
    \centering
    \resizebox{0.65\columnwidth}{!}{\includegraphics[]{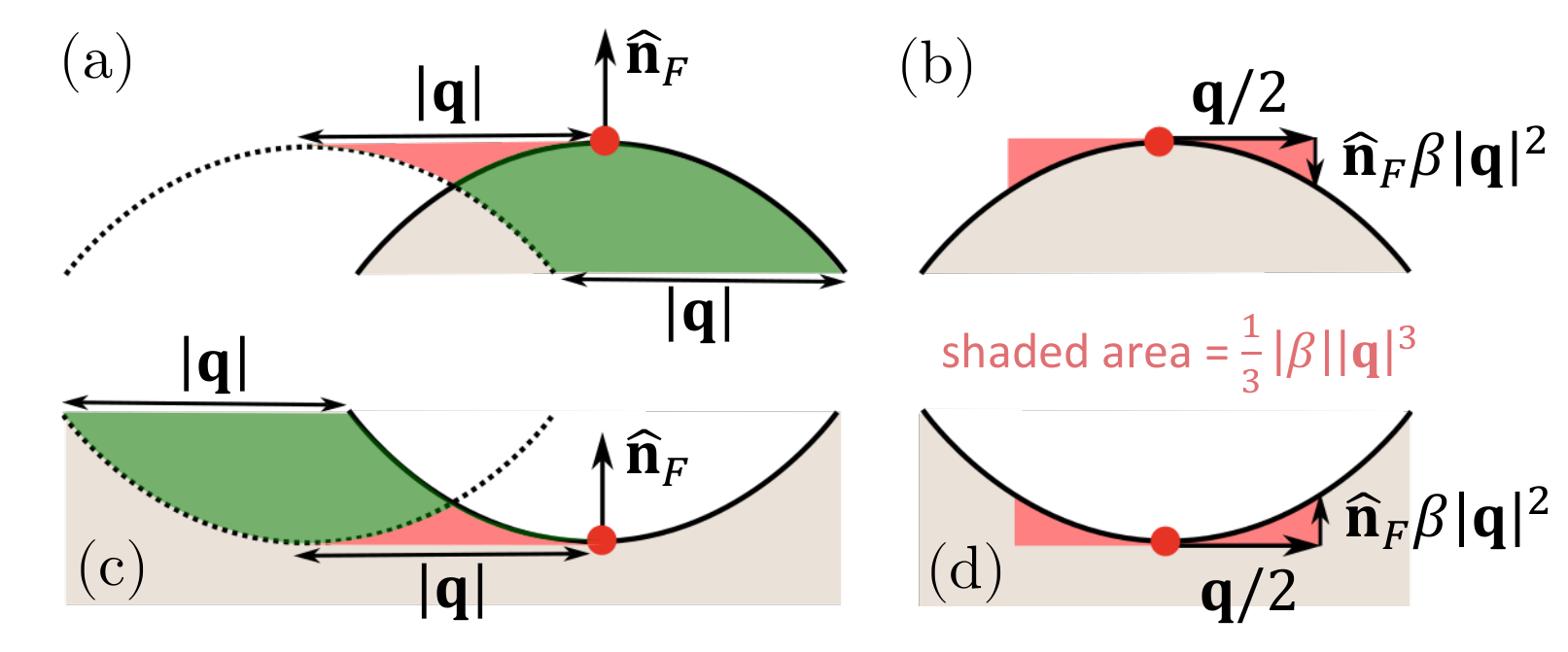}}
    \caption{Geometric interpretation of $S(\bq)$. (a) Around a convex Fermi surface (which is shown as the solid black curve, with its $\bq$-shifted counterpart represented as the dashed curve), the $S^{(1)}(\bq)$ term in $S(\bq)$ (see Eq. \eqref{supp_eq: alpha}) corresponds to the entire shaded region including both the green and red parts, while $S(\bq)$ only corresponds to the green shaded part. (b) The red shaded region around a Fermi surface critical point (with $\hat{\bq}\cdot \hat{\bn}_F =0 $ and labeled as the red dot) corresponds to an over-counted area on this $\hat{\bn}_F$-$\hat{\bq}$ plane, which should be subtracted from $S^{(1)}(\bq)$ to give the correct $S(\bq)$. To leading order, by expanding the Fermi surface quadratically around the critical point, this over-counted area is $\frac{1}{3}\abs{\beta}\abs{\bq}^3$, where $\beta$ is defined in the figure such that $\frac{1}{2}\bq+\beta\abs{\bq}^2\hat{\bn}_F$ is on the Fermi surface (with the origin here, labeled as the red dot, being the critical point). (c) and (d) show a similar analysis for a concave Fermi surface, which gives the same over-counted area. }
    \label{fig: overcounted area}
\end{figure}
As alluded to in the main text, $\int_\bk f_\bk (1-f_{\bk+\bq})$ has a simple geometrical meaning in terms of the difference between the volume of the Fermi sea and the overlapping volume of the Fermi sea with its counterpart shifted by $\bq$. Our task here is to evaluate this volume difference to order $\abs{\bq}^3$. To that end, we parametrize each sheet of the Fermi surface $\bk_F(\bs)$ by $\bs=\{s^\mu, \mu=1,2\}$, with the induced metric from the 3D Euclidean reciprocal space ($\bk=k_a\bee^a$) defined as follows and satisfies
\begin{equation}\label{supp_eq: def of metric g}
    g_{\mu\nu}(\bs) = \partial_\mu \bk_F(\bs)\cdot \partial_\nu \bk_F(\bs),\quad g\equiv \det g = \frac{1}{2}\epsilon^{\mu\sigma}\epsilon^{\nu\tau} g_{\mu\nu}g_{\sigma\tau},\quad g^{\mu\nu}\partial_\mu k_{Fa}\partial_\nu k_{Fb}  = \delta_{ab}-\delta_{ac}\delta_{bd}\hat{n}^c_F\hat{n}^d_F,
\end{equation}
where $\partial_\mu \equiv \frac{\partial}{\partial s^\mu}$, $g^{\mu\nu} = \frac{1}{g}\epsilon^{\mu\sigma}\epsilon^{\nu\lambda}g_{\sigma\lambda}$ is the inverse induced metric, $\delta_{ab} = \bee_a \cdot \bee_b$ is the Euclidean metric of the flat 3D space, and $\epsilon^{\mu\nu}$ is the 2D Levi-Civita symbol. As introduced before, $\hat{\bn}_F(\bs)$ is the unit outward normal vector on the Fermi surface, with $\hat{n}_F^a(\bs)\partial_\mu k_{Fa}(\bs) = 0$ and $\delta_{ab}\hat{n}_F^a(\bs)\hat{n}_F^b(\bs)=1$. The orientation of the 2D Levi-Civita symbols $\epsilon^{\mu\nu}$ and $\eps_{\mu\nu}$ (with $\eps^{\mu\nu} \eps_{\sigma\tau}  = \delta^\mu_\sigma \delta^\nu_\tau - \delta^\nu_\sigma \delta^\mu_\tau$) is fixed by 
\begin{equation}
    dA_F = \hat{\bn}_F \cdot (\partial_\mu\bk_F \times \partial_\nu \bk_F) ds_1^\mu ds_2^\nu = \sqrt{g}\epsilon_{\mu\nu} ds_1^\mu ds_2^\nu > 0\;.
\end{equation}
From this, it can be seen that
\begin{equation}\label{supp_eq: relation between partial k and nF}
    \epsilon^{abc} \partial_\mu  k_{Fa} \partial_\nu k_{Fb} = \sqrt{g} \epsilon_{\mu\nu}\hat{n}^c_F \quad \text{and} \quad \epsilon^{\mu\nu} \partial_\mu  k_{Fa} \partial_\nu k_{Fb} = \sqrt{g}\epsilon_{abc} \hat{n}^c_F\;.
\end{equation}
Throughout this work, only repeated upper and lower indices are contracted. \\

To capture the effect of Fermi surface curvature, which is essential for the $\abs{\bq}^3$-term in $S(\bq)$, we introduce the extrinsic curvature tensor (also known as the second fundamental form \cite{frankel2004geometry}) on the Fermi surface
\begin{equation} \label{supp_eq: def of K}
    K_{\mu\nu}(\bs) \equiv \hat{\bn}_F(\bs)\cdot \partial_\mu \partial_\nu \bk_F(\bs) = -\partial_\mu \hat{\bn}_F\cdot\partial_\nu\bk_F = -\partial_\nu \hat{\bn}_F\cdot\partial_\mu\bk_F.
\end{equation}
It is related to the two principal curvatures $\kappa_{1,2}$ via the following eigenvalue equation:
\begin{equation}
    K_{\mu\nu} u^\nu _i = \kappa_i g_{\mu\nu} u^\nu_i\;\;\;\;(i=1,2), \quad g_{\mu\nu}u^\mu_i u^\nu_j=\delta_{ij}, \quad \sum_{i=1}^2 u^\mu_i u^\nu_i = g^{\mu \nu}, 
\end{equation}
where $u_i$ are the two principal directions. This allows $K_{\mu\nu}$ to be expressed as
\begin{equation}\label{supp_eq: relating K to curvature}
    K_{\mu\nu} = \sum_{i=1}^2 g_{\mu\sigma} u^\sigma_i \kappa_i u^\tau _i g_{\tau \nu} \implies \tr K \equiv K_{\mu\nu} g^{\mu\nu} = \kappa_1+\kappa_2, \;\;\tr K^2\equiv K_{\mu\nu} g^{\nu\sigma} K_{\sigma\tau} g^{\tau\mu} = \kappa_1^2+\kappa_2^2.
\end{equation}
Since both $\hat{\bn}_F \cdot \partial_\mu \hat{\bn}_F =0$ and $\hat{\bn}_F\cdot\partial_\mu \bk_F=0$, we can expand $\partial_\mu \hat{\bn}_F$ as $\partial_\mu \hat{\bn}_F = -K_{\mu\nu} g^{\nu\sigma}\partial_\sigma\bk_F$, from which we can obtain
\begin{equation}
    \hat{\bn}_F \cdot(\partial_\mu \bn_F \times \partial_\nu \bn_F) = \epsilon_{\mu\nu} \frac{\det K}{\sqrt{g}} = \epsilon_{\mu\nu} \sqrt{g} \kappa_1 \kappa_2.
\end{equation}
From the Gauss-Bonnet theorem, we recognize that the Euler characteristic of Fermi surface can be interpreted as a winding number:
\begin{equation}
    \chi_{\rm FS} =  \frac{1}{2\pi} \int_{\rm FS} dA_F\; \kappa_1\kappa_2= \frac{1}{2\pi} \int ds_1^\mu ds_2^\nu\;\hat{\bn}_F \cdot(\partial_\mu \bn_F \times \partial_\nu \bn_F).
\end{equation}
\subsubsection{Volume calculation}
From Eq. \eqref{supp_eq: alpha}, and by inspecting Fig. \ref{fig: overcounted area}, we can see that $S^{(1)}(\bq)$ has \textit{over-counted} $\bk$-space volume around Fermi surface critical lines where $\hat{\bn}_F \cdot \hat{\bq}=0$. As we see from Fig. \ref{fig: overcounted area}, this over-counted volume has a leading contribution of order $\abs{\bq}^3$, corresponding to the $S^{(3)}_{\rm G}(\bq)$ term in $S(\bq)$, which we evaluate below. Let us consider the Fermi surface locally around each critical point $\bk_F(\bs_0)$ (with $\hat{\bn}_F(\bs_0)\cdot \hat{\bq}=0$) to obtain its quadratic profile in the $\bk$-space plane spanned by $\hat{\bq}$ and $\hat{\bn}_F$. In particular, we want to obtain $\beta$ as indicated in Fig. \ref{fig: overcounted area}, such that the over-counted area on this plane is $\frac{1}{3}\abs{\beta}\abs{\bq}^3$. We can find $\beta$ by moving away from $\bs_0$ by an amount $\delta\bs$ that satisfies
\begin{equation}\label{supp_eq: beta in terms of s}
    \partial_\mu \bk_F \delta s^\mu + \frac{1}{2}\partial_\mu \partial_\nu \bk_F \delta s^\mu \delta s^\nu = \frac{1}{2} \bq + \beta \abs{\bq}^2 \hat{\bn}_F \implies \beta \abs{\bq}^2 = \frac{1}{2} K_{\mu\nu} \delta s^\mu \delta s^\nu \quad \text{and}\quad  \frac{1}{2}\abs{\bq} = \hat{\bq}\cdot\partial_\mu \bk_F \delta s^\mu
\end{equation}
As we are staying on the plane defined by $\hat{\bq}$ and $\hat{\bn}_F$, hence orthogonal to $\hat{\bt} \equiv \hat{\bn}_F \times \hat{\bq}$, we have $\hat{\bt}\cdot \partial_\mu\bk_F \delta s^\mu = 0$ which allows for a parametrization as $\delta s^\mu = \mathfrak{s}\eps^{\mu\nu}\hat{\bt}\cdot\partial_\nu \bk_F$. The proportionality factor $\mathfrak{s}$ is determined by requiring $\frac{1}{2}\abs{\bq} = \hat{\bq}\cdot\partial_\mu \bk_F \delta s^\mu$, which fixes $\mathfrak{s}$ (upon using Eq. \eqref{supp_eq: relation between partial k and nF}):
\begin{equation}
    \frac{1}{2}\abs{\bq} = \mathfrak{s}\hat{\bq}\cdot\partial_\mu \bk_F \eps^{\mu\nu}\hat{\bt}\cdot\partial_\nu \bk_F = \mathfrak{s}\sqrt{g}\;\hat{\bt}\cdot (\hat{\bn}_F\times \hat{\bq})  \implies \mathfrak{s} = \frac{\abs{\bq}}{2\sqrt{g}}.
\end{equation}
Substituting back into Eq. \eqref{supp_eq: beta in terms of s}, we obtain
\begin{equation}
    \beta = \frac{1}{8g} K_{\mu\nu} \eps^{\mu\sigma}\epsilon^{\nu\tau} (\hat{\bt}\cdot \partial_\sigma \bk_F) (\hat{\bt}\cdot \partial_\tau \bk_F), \;\;\;\hat{\bt}\equiv\hat{\bn}_F \times \hat{\bq}.
\end{equation}
Notice that this expression is independent of the choice of parameterization for the Fermi surface.

To obtain the over-counted volume, we need to integrate the area element found above along critical lines defined by $\hat{\bq}\cdot \hat{\bn}_F (\bs)=0$. As we will show, $S^{(3)}_{\rm G}(\bq) = S^{(3)}_{\rm G}(\hat{\bq})\abs{\bq}^3$ can be expressed as a Fermi-surface integral. To achieve that, we set up a local Fermi surface tangent-frame coordinate system $(u,v)$ at each critical point $\bk_F(\bs_0)$ (with $\bs_0$ corresponding to $(u,v)=(0,0)$), such that $u$ parametrizes the direction parallel ($\parallel$) to the critical line $\hat{\bq}\cdot \hat{\bn}_F (\bs)=0$, and $v$ parametrizes the direction perpendicular ($\perp$) to the critical line. Thus, 
\begin{equation}\label{supp_eq: local tangent frame}
    \begin{cases}
         ds^\mu_\parallel = \epsilon^{\mu\nu} \partial_\nu(\hat{\bq}\cdot\hat{\bn}_F) du\\
         d s^\mu_\perp = g^{\mu\nu} \partial_\nu(\hat{\bq}\cdot\hat{\bn}_F) dv
    \end{cases},
    \;\;\; dA_F = \sqrt{g} g^{\mu\nu} \partial_\mu(\hat{\bq}\cdot\hat{\bn}_F)\partial_\nu(\hat{\bq}\cdot\hat{\bn}_F) du dv,\;\;\; \hat{\bq}\cdot \hat{\bn}_F = g^{\mu\nu} \partial_\mu(\hat{\bq}\cdot\hat{\bn}_F)\partial_\nu(\hat{\bq}\cdot\hat{\bn}_F) v
\end{equation}
Integrating along the critical line (with the line element $du >0$), we obtain
\begin{equation}\label{supp_eq: intermediate gamma}
    S^{(3)}_{\rm G}(\hat{\bq}) = \frac{1}{(2\pi)^3} \oint du\;\frac{1}{3} \abs{\beta \epsilon^{\mu\nu}(\hat{\bt}\cdot \partial_\mu \bk_F) (\hat{\bq}\cdot\partial_\nu \hat{\bn}_F)}
\end{equation}

\subsubsection{Result}
Notice that, using Eqs. \eqref{supp_eq: def of metric g}, \eqref{supp_eq: relation between partial k and nF} and \eqref{supp_eq: def of K}, we have the following identities
\begin{subequations}
\begin{align}
    K_{\mu\nu} \eps^{\mu\sigma}\epsilon^{\nu\tau} (\hat{\bt}\cdot \partial_\sigma \bk_F) (\hat{\bt}\cdot \partial_\tau \bk_F) &= \sqrt{g} \epsilon^{\mu\nu} (\hat{\bt}\cdot \partial_\mu \bk_F) (\hat{\bq}\cdot\partial_\nu \hat{\bn}_F) \\
K_{\mu\nu} \eps^{\mu\sigma}\epsilon^{\nu\tau} (\hat{\bt}\cdot \partial_\sigma \bk_F) (\hat{\bt}\cdot \partial_\tau \bk_F) &= g K_{\mu\nu} g^{\mu\sigma} g^{\nu\tau} (\hat{\bq}\cdot \partial_\sigma \bk_F)(\hat{\bq}\cdot \partial_\tau \bk_F)
\end{align}
\end{subequations}
The absolute magnitude sign in the integrand of Eq. \eqref{supp_eq: intermediate gamma} is thus redundant, and we can rewrite $S^{(3)}_{\rm G}(\hat{\bq})$ as follows:
\begin{equation}\label{supp_eq: critical line integral and def of I}
    S^{(3)}_{\rm G}(\hat{\bq}) = \frac{1}{192\pi^3} \oint du\sqrt{g}\; I(\bs(u), \hat{\bq}), \quad I(\bs,\hat{\bq}) = [K_{\mu\nu} g^{\mu\sigma} g^{\nu\tau} (\hat{\bq}\cdot \partial_\sigma \bk_F)(\hat{\bq}\cdot \partial_\tau \bk_F)]^2.
\end{equation}
Finally, we can rewrite the above as a Fermi surface integral using Eq. \eqref{supp_eq: local tangent frame}:
\begin{equation}\label{supp_eq: gamma G}
    S^{(3)}_{\rm G}(\hat{\bq}) = \frac{1}{192\pi^3} \sum_m \int_{{\rm FS}_m} dA_F(\bs) \;\delta(\hat{\bq}\cdot\hat{\bn}_F)\;I(\bs, \hat{\bq})\;.
\end{equation}
This gives Eq.~\eqref{eq: gamma_G 2} in the main text.

\subsubsection{Benchmark: ellipsoidal Fermi surface}
As a sanity check, let us benchmark Eq.~\eqref{eq: gamma_G 2} in the case of an ellipsoidal Fermi surface. Without loss of generality, the Fermi surface is taken as $\bk_{F,a} \eta^{ab}\bk_{F,b} = k_F^2$, with $\eta = \text{diag}(\eta_1,\eta_2,\eta_3)$ and $\det\eta=1$. The ellipsoidal Fermi surface can be parameterized by polar coordinates, $\bs=(\theta,\phi)$. Defining $\hat{\boldsymbol{\tau}}(\theta, \phi) = (\sin\theta\cos\phi, \sin\theta\sin\phi, \cos\theta)^T$, we have
\begin{equation}\label{supp_eq: ellipsoid kF and nF}
    \bk_F(\theta, \phi) = k_F\eta^{-1/2}\hat{\boldsymbol{\tau}}(\theta,\phi), \quad \hat{\bn}_F(\theta, \phi) =\frac{1}{\sqrt{Q(\theta, \phi)}} \eta^{1/2}\;\hat{\boldsymbol{\tau}}(\theta, \phi), \quad Q(\theta, \phi) \equiv \hat{\boldsymbol{\tau}}^T \eta \hat{\boldsymbol{\tau}}.
\end{equation}
As for the metric tensor and the curvature tensor, we have
\begin{equation}\label{supp_eq: ellipsoid g and K}
    g_{\mu\nu}(\theta, \phi) = k_F^2\partial_\mu\hat{\boldsymbol{\tau}}^T\eta^{-1}\partial_\nu \hat{\boldsymbol{\tau}} \quad\text{and}\quad K_{\mu\nu}(\theta, \phi) = \frac{k_F}{\sqrt{Q}}\hat{\boldsymbol{\tau}}^T\partial_\mu\partial_\nu\hat{\boldsymbol{\tau}}.
\end{equation}
The surface integral measure in Eq.~\eqref{supp_eq: gamma G}  can then be expressed as $dA_F = d\theta d\phi \sqrt{\det g} = d\theta d\phi k_F^2\sin\theta\sqrt{Q(\theta, \phi)} $. Notice that $\hat{\bq}\cdot\hat{\bn}_F(\theta, \phi) = 0 $ implies 
\begin{equation}\label{supp_eq: critical line position, theta}
    \tan \theta = -\frac{\sqrt{\eta_3}\cot\theta_\bq}{\sqrt{\eta_1}\cos\phi\cos\phi_\bq+\sqrt{\eta_2}\sin\phi\sin\phi_\bq},
\end{equation}
where $\hat{\bq} \equiv (\sin\theta_\bq\cos\phi_\bq, \sin\theta_\bq\sin\phi_\bq, \cos\theta_\bq)$. With this we can easily integrate out $\delta(\hat{\bq}\cdot\hat{\bn}_F)$ and obtain from Eq.~\eqref{supp_eq: gamma G}
\begin{equation}\label{supp_eq: ellipsoid gamma}
    S^{(3)}_{\rm G}(\hat{\bq}) = \frac{1}{192\pi^3} \int_0^{2\pi} d\phi \frac{\sin\theta^* \sqrt{Q(\theta^*,\phi)}}{\abs{\hat{\bq}\cdot\partial_\theta \hat{\bn}_F}_{\theta=\theta^*}} k_F^2 I(\theta^*,\phi, \hat{\bq}),
\end{equation}
where $\theta^*$ is obtained from Eq.~\eqref{supp_eq: critical line position, theta}, for any given $\phi$ and $\hat{\bq}$, and $I(\theta^*,\phi, \hat{\bq})$ is obtained from its definition in Eq.~\eqref{supp_eq: critical line integral and def of I} together with the expressions in Eq.~\eqref{supp_eq: ellipsoid kF and nF} and ~\eqref{supp_eq: ellipsoid g and K}. Note that taking into account factors of $k_F$ from $I$, $k_F$ eventually drops out of the expression for $S^{(3)}_{\rm G}(\hat{\bq})$, as it should for this is a dimensionless quantity. While the integral in Eq.~\eqref{supp_eq: ellipsoid gamma} may not be evaluated analytically for a generic ellipsoid, it can be easily computed numerically. 

This result can be matched with an alternative, analytically exact calculation for the one-band case with an ellipsoid Fermi surface. Recall the geometric meaning of $S(\bq)$ (in the one-band case) as the difference between the volume of Fermi sea and the overlapping volume of two identical but shifted (by amount $\bq$) Fermi seas. From the overlapping volume of ellipsoids, we obtain
\begin{equation}
    S(\bq) = \frac{1}{(2\pi)^3} \left[\pi k_F^2\sqrt{\bq^T \eta\bq} - \frac{\pi}{12}(\bq^T \eta\bq)^{3/2}\right] \implies S^{(3)}(\hat{\bq}) = \frac{1}{96\pi^2}\left[\eta_1\sin^2\theta_\bq\cos^2\phi_\bq+ \eta_2\sin^2\theta_\bq\sin^2\phi_\bq+\eta_3 \cos^2\theta_\bq \right]^{3/2}.
\end{equation}
By direct computations, we have checked that this indeed matches with Eq.~\eqref{supp_eq: ellipsoid gamma}.

\subsection{Quantum geometric contribution: $S^{(3)}_{\rm QG}$}\label{supp_sec: quantum geometric}

We now consider quantum geometric contributions to Eq. \eqref{supp_eq: symmetrized Sq} by expanding $\mathscr{P}_{m,n}(\bk, \bq)\equiv \Tr[P_m(\bk) P_n(\bk+\bq)]$ to higher orders in $\bq$. First of all, the first-order term vanishes identically:
\begin{equation}
\begin{split}
    q_a \Tr[P_m(\bk)\partial^a P_n(\bk)] &= q^a \sum_{\sigma,\sigma'} \left(U^\dagger_{m, \sigma}(\bk)\partial_a U_{\sigma, n}(\bk) U^\dagger_{n, \sigma'}(\bk)U_{\sigma',m}(\bk)+ U^\dagger_{m, \sigma}(\bk)U_{\sigma, n}(\bk) \partial_a U^\dagger_{n, \sigma'}(\bk)U_{\sigma',m}(\bk)\right)\\
    &=q^a  \left(\sum_\sigma U^\dagger_{m, \sigma}(\bk)\partial_a U_{\sigma, n}(\bk) + \sum_{\sigma'} \partial_a U^\dagger_{n, \sigma'}(\bk)U_{\sigma',m}(\bk)\right) \delta_{m,n} \\
    & = q^a \partial_a \left(\sum_\sigma U^\dagger_{m,\sigma}(\bk)U_{\sigma,m}(\bk)\right)\delta_{m,n} = 0.
\end{split}
\end{equation}
The same conclusion can be reached using the continuum formulation with Eq.~\eqref{supp_eq: continuum formulation for projector}. Thus the non-analytic $\abs{\bq}^3$-contribution to $S(\bq)$ comes from the $q^2$ term in expanding $\mathscr{P}_{m,n}(\bk, \bq)$. \\

Let us now make an important observation that only the $m=n$ terms in Eq.~\eqref{supp_eq: symmetrized Sq} can provide singular contributions, under our operating assumption that different components of Fermi surfaces are separated and do not intersect. Recall from Eq.~\eqref{supp_eq: symmetrized Sq}:
\begin{equation}
\begin{split}
    S(\bq) = \frac{1}{2}\int_{\rm BZ} \frac{d^3\bk}{(2\pi)^3} \sum_{m,n} \mathscr{P}_{m,n}(\bk, \bq)(f_{m,\bk}-f_{n,\bk+\bq}) \sgn(E_{n,\bk+\bq}-E_{m,\bk}).
\end{split}
\end{equation}
For $m \neq n$, it is guaranteed that in each of the connected region in the $\bk$-space where the above integrand is non-vanishing (i.e., $f_{m,\bk}-f_{n,\bk+\bq} \neq 0$), $\sgn(E_{n,\bk+\bq}-E_{m,\bk})$ is fixed by the energy ordering of the bands in that region and thus only contribute an overall sign that is insensitive to $\bq$. Hence, upon shifting the momentum variable ($\bk+\bq \rightarrow \bk$) for the term involving $f_{n,\bk+\bq}$, the entire integrand in Eq.~\eqref{supp_eq: symmetrized Sq} is manifestly analytic in $\bq$. It is important that the Fermi surface of $m$ and that of $n$ are separated, so that for small $\bq$ the shifting of the integration domain would not lead to any intersection between the domain boundary and Fermi surfaces, which ensures the analyticity in $\bq$.\\

We thus focus on $m=n$ in Eq.~\eqref{supp_eq: symmetrized Sq}. At order $q^2$, $\mathscr{P}_{m,n}(\bk, \bq)\equiv \Tr[P_m(\bk) P_m(\bk+\bq)]$ gives $-q_aq_b\mathcal{G}^{ab}_m(\bk)$, where ($\partial^a \equiv \frac{\partial}{\partial k_a}$)
\begin{equation}
    \mathcal{G}^{ab}_m(\bk) = \frac{1}{2}\Tr[\partial^aP_m(\bk) \partial^bP_m(\bk)]
\end{equation}
is identified as the quantum metric for the $m$-th band \cite{provost1980riemannian, MarzariVanderbilt1997, Roy2014}.
Notice that the same conclusion holds also for the continuum formulation using Eq.~\eqref{supp_eq: continuum formulation for projector}, which gives $\mathcal{G}^{ab}_m(\bk) =\braket{u_{m,\bk}|\partial^a u_{m,\bk}}\braket{u_{m,\bk}|\partial^b u_{m,\bk}}+\frac{1}{2}\left[\braket{\partial^a u_{m,\bk}|\partial^b u_{m,\bk}}+(a\leftrightarrow b)\right]$. When there is no ambiguity, we would sometime suppress the $\bk$-dependence for notational simplicity. Expressing the rest to linear order in $q$: $(f_{m,\bk}-f_{m,\bk+\bq}) \sgn(E_{m,\bk+\bq}-E_{m,\bk}) = \abs{\bq\cdot \bv_{m,\bk}}\delta(E_F-E_{m,\bk})$, we obtain the following quantum geometric contribution:
\begin{equation}\label{supp_eq: gamma QG}
    S^{(3)}_{\rm QG}(\bq) = \frac{1}{16\pi^3}\sum_m \int_{\rm{FS}_m} dA_F \;\mathcal{G}^{ab}_m q_a q_b\abs{\hat{\bn}_F\cdot \bq}
\end{equation}
where $\int_{\rm{FS}_m} dA_F $ represents the surface integral over the $m$-th Fermi surface and $\hat{\bn}_F$ is the corresponding outward unit normal vector. This gives Eq.~\eqref{eq: gamma QG as FS integral} in the main text.

\section{Bipartite charge fluctuation}\label{supp_sec: bipartite fluctuation}
\setcounter{equation}{0}
\setcounter{figure}{0} 

In this section we derive the logarithmic coefficient for the bipartite charge fluctuation finite-size scaling in 3D Fermi gas with an ellipsoid partition surface in the real space, making use of Eq.~\eqref{supp_eq: gamma G} and ~\eqref{supp_eq: gamma QG}. First, we establish the precise connection between charge fluctuations and the structure factor. The bipartite charge fluctuation is defined as
\begin{equation}
\begin{split}
    \cc{Q^2_A} \equiv \langle Q^2_A\rangle - \langle Q_A\rangle^2 &= \int_A d^3\br \int_Ad^3\br'\;\cc{\rho(\br)\rho(\br')}\\
    &= \int_A d^3\br \int_Ad^3\br'\int\frac{d^3\bq}{(2\pi)^3}\int\frac{d^3\bq'}{(2\pi)^3} e^{i\bq\cdot\br+i\bq'\cdot\br'}\cc{\rho_\bq\rho_{\bq'}}
    \label{eq: q2-general}
\end{split}
\end{equation}
Here, $Q_A = \int_A d^3\br \rho(\br)$ is the charge in region $A$, with $\rho(\br)= \int \frac{d^3\bq}{(2\pi)^3} e^{i\bq\cdot\br}\rho_\bq$ the real-space density operator, and $\cc{\rho_1 \rho_2} \equiv \langle \rho_1\rho_2\rangle - \langle\rho_1\rangle \langle\rho_2\rangle$ is the connected two-point correlation function. Notice that the $\bq$-integrals are carried out over the entire (unbounded) momentum space. In our case of interest with lattice translational symmetry, using the form in Eq.~\eqref{supp_eq: general rho expression}, it can be seen from Wick's theorem that $\cc{\rho_\bq\rho_{\bq'}} = 0$ if $\bq' \neq -\bq$ (mod $\bG$), where $\bG$ is a reciprocal lattice vector. Therefore, we can express the charge fluctuations as
\begin{equation}\label{supp_eq: general expression for charge fluctuations}
\begin{split}
    \cc{Q^2_A} &= \int_A d^3\br \int_Ad^3\br'\int\frac{d^3\bq}{(2\pi)^3}\sum_{\bG} e^{i\bq\cdot\br+i(-\bq+\bG)\cdot\br'}\frac{\cc{\rho_\bq\rho_{-\bq+\bG}}}{V}\\
    &=\sum_{\bG}\int\frac{d^3\bq}{(2\pi)^3} \mathcal{F}_A(\bq)\mathcal{F}_A(-\bq+\bG)\frac{\cc{\rho_\bq\rho_{-\bq+\bG}}}{V}\equiv \sum_\bG \mathscr{Q}_A^{(\bG)}
\end{split}
\end{equation}
where $\mathcal{F}_A(\bq)\equiv \int_A d^3\br e^{i\bq\cdot\br} $ is the spatial form factor for the partition region $A$. It is straightforward to obtain $\mathcal{F}_A(\bq)$ when the partition surface $\partial A$ is a sphere of radius $L$:
\begin{equation}\label{supp_eq: spatial form factor for sphere}
    \mathcal{F}_A(\bq) = \frac{4\pi}{\abs{\bq}^3}[\sin(\abs{\bq}L)-\abs{\bq}L \cos(\abs{\bq}L)].
\end{equation}

Notice that the $\bG=0$ term in Eq.~\eqref{supp_eq: general expression for charge fluctuations} is determined by the structure factor, with $\mathscr{Q}_A^{(0)} = \int \frac{d^3\bq}{(2\pi)^3}\abs{\mathcal{F}_A(\bq)}^2 S(\bq)$. For a system with continuous translational symmetry, this is all that contributes to the charge fluctuations. However for generic lattice systems, $\bG\neq 0 $ terms contribute as well. Nonetheless, as far as the \textit{logarithmic} finite-size scaling is concerned, only the $\bG=0$ term (and hence determined by $S(\bq)$) matters, as we explain below.

\subsubsection{Justification for focusing on $\bG=0$}
Using Eq.~\eqref{supp_eq: spatial form factor for sphere}, we can see that the evaluation of $\mathscr{Q}_A^{(\bG)}$ involves the following momentum integral
\begin{equation}
    \int_0^{a^{-1}} d\abs{\bq} \frac{\cc{\rho_\bq\rho_{-\bq+\bG}}}{\abs{\bq}\abs{\bq-\bG}^3}[\sin(\abs{\bq}L)-\abs{\bq}L\cos(\abs{\bq}L)]
     \cdot[\sin(\abs{\bq-\bG}L)-\abs{\bq-\bG}L\cos(\abs{\bq-\bG}L)]
\end{equation}
where we have introduced an ultra-violet cutoff $a^{-1}$ for the momentum integral. Expanding $\cc{\rho_\bq\rho_{-\bq+\bG}}$ in power series of $\abs{\bq}$, we first notice that the zeroth order term vanishes due to total charge conservation. Plugging in subsequent higher order terms ($\abs{\bq}$, $\abs{\bq}^2$, etc. ) into the above integral, we observe that no logarithmically divergent terms (such as $\log(L/a)$ or $L^2\log(L/a)$) can possibly arise whenever $\bG \neq 0$. Logarithmic type divergence arises for $\bG=0$, in which case we have
\begin{equation}
    \int_0^{a^{-1}} d\abs{\bq} \abs{\bq}^{\delta-4}[\sin(\abs{\bq}L)-\abs{\bq}L\cos(\abs{\bq}L)]^2,
\end{equation}
where $\abs{\bq}^\delta$ is extracted from the expansion of $\cc{\rho_\bq\rho_{-\bq}}$. When $\delta=1$, we obtain a log-type divergence of the form $L^2\log(L/a)$. When $\delta=3$, we obtain the $\log(L/a)$ divergence, which is the main interest of this work. While the above argument is based on the specific real-space form factor $\mathcal{F}_A(\bq)$ for a spherical partition, it is straightforward to extend the conclusion to generic ellipsoidal partitions by a uniform coordinate transformation (detailed below). For our purpose of characterizing the \textit{logarithmic} finite-size scaling, we shall thus focus on the $\bG=0$ term in Eq.~\eqref{supp_eq: general expression for charge fluctuations} for the rest of this section. Next, we set up the stage for evaluating the logarithmic coefficient.

\subsubsection{General ellipsoidal partition}
The real-space subregion $A$ is defined by the interior of the partition surface. We take $\partial A$ to be an ellipsoid in the Euclidean real space specified by a quadratic form $R_{ab}$:
\begin{equation}\label{supp_eq: def of the real space partition surface}
    \partial A:\;\;\; r^aR_{ab}r^b =  L^2,\;\;\text{with}\;\;\det R=1.
\end{equation}
The form factor $\mathcal{F}_A(\bq)\equiv \int_A d^3\br e^{i\bq\cdot\br} $ can be obtained via the following change of variables:
\begin{equation}
    r^a = \tilde{r}^\alpha v^a_\alpha, \quad \tilde{q}_\alpha=q_av^a_\alpha, \quad \text{with}\;\;v^a_\alpha R_{ab}v^b_\beta = \delta_{\alpha\beta}, 
\end{equation}
such that
\begin{equation}
    \mathcal{F}_A(\bq) = \int d^3\tilde{\br}\;\theta(L^2-\abs{\tilde{r}}^2) e^{i\tilde{\bq}\cdot\tilde{\br}} = \frac{4\pi}{\abs{\tilde{\bq}}^3}\left[\sin(\abs{\tilde{\bq}}L)-\abs{\tilde{\bq}}L\cos(\abs{\tilde{\bq}}L)\right] \equiv \mathcal{F}(\abs{\tilde{\bq}}, L).
\end{equation}
As far as the logarithmic finite-size scaling behavior is concerned, following Eq.~\eqref{supp_eq: general expression for charge fluctuations} and the justification provided above, we have
\begin{equation}\label{supp_eq: starting point for charge fluctuation}
    \cc{Q^2_A} = \frac{1}{(2\pi)^3}\int d^2\hat\Omega(\tilde{\bq}) \int d\abs{\tilde{\bq}}\mathcal{F}(\abs{\tilde{\bq}},L)^2\abs{\tilde{\bq}}^2 S(\bq).
\end{equation}
This serves as the starting point for our study of finite-size scaling. As we will see, upon substituting in the expansion in Eq.~\eqref{supp_eq: expansion of Sq}, as a function of the linear system size $L$ of the region $A$, the finite-size scaling generally contains a $\log L$-term whose coefficient features an interesting interplay between the geometry of the real-space partition surface and the geometry (as well as quantum geometry) of the Fermi surface. To warm up, let us first derive the leading divergent term of charge fluctuation from Eq.~\eqref{supp_eq: alpha}, which is related to the well-known Gioev-Klich-Widom formula for the leading divergence of entanglement entropy in Fermi gases \cite{GioveKlich2006}. 

\subsection{$L^2\log L$-scaling and Fermi surface area}
Substituting in the leading $S^{(1)}(\hat{\bq})\abs{\bq}$ term in Eq.~\eqref{supp_eq: expansion of Sq} into Eq.~\eqref{supp_eq: starting point for charge fluctuation}, we obtain the leading divergence from
\begin{equation}
    \int_0^{a^{-1}} d\abs{\tilde{\bq}}\mathcal{F}(\abs{\tilde{\bq}},L)^2\abs{\tilde{\bq}}^3 = 8\pi^2 L^2\log(L/a)+ \;...,
\end{equation}
where $a$ is an ultra-violet cutoff at the lattice length scale. The leading divergence in $\cc{Q^2_A}$ is thus
\begin{equation}
\begin{split}
    \cc{Q^2_A} &= \left(\frac{1}{\pi}\int d^2\hat{\Omega}(\tilde{\bq})\;S^{(1)}(\hat{\bq})\frac{\abs{\bq}}{\abs{\tilde{\bq}}}\right) L^2\log L \;+\; ...\;,\\
    &=\left( \frac{1}{16\pi^4}\sum_m\int_{{\rm FS}_m}dA_F \int d^2\hat{\Omega}(\tilde{\bq}) \abs{\hat{\bn}_F\cdot \hat{\bq}}\frac{\abs{\bq}}{\abs{\tilde{\bq}}}\right) L^2\log L \;+\; ...\;.
\end{split}
\end{equation}
For each point on the Fermi surface we perform the following angular integral:
\begin{equation}\label{supp_eq: angular integral warmup}
\begin{split}
    \int d^2\hat{\Omega}(\tilde{\bq}) \abs{\hat{\bn}_F\cdot \hat{\bq}}\frac{\abs{\bq}}{\abs{\tilde{\bq}}} &= \int d^3\tilde{\bq}\;\delta(\abs{\tilde{\bq}}-1)  \abs{\hat{\bn}_F\cdot\bq} =\int d^3\bq\;\delta\left(\sqrt{q_a (R^{-1})^{ab}q_b}-1\right)\abs{\hat{\bn}_F\cdot\bq}\\
    &=\int d^3\bx \;\delta\left(\sqrt{x^aR_{ab}x^b}-1\right)\abs{\hat{n}_F^aR_{ab}x^b}\\
    &=\int'_{\partial A} dA_R\;\abs{\hat{\bn}_F\cdot\hat{\bn}_R}.
\end{split}
\end{equation}
In the second line above we made a change of variable: $q_a = R_{ab}x^b$, and in the third line the surface integral ($\int'$) is performed over the \textit{normalized} partition surface (i.e., over $x^aR_{ab}x^b=1$ instead of Eq.~\eqref{supp_eq: def of the real space partition surface}), with $\hat{\bn}_R$ the unit outward normal vector on the partition surface in real space. Substituting this back into the full expression for the charge fluctuation, we obtain
\begin{equation}
    \cc{Q^2_A} = \left( \frac{1}{16\pi^4}\sum_m\int_{{\rm FS}_m}dA_F \int_{\partial A}dA_R \abs{\hat{\bn}_F\cdot\hat{\bn}_R}\right)\log L + \;...\;,
\end{equation}
where the $L^2$ factor has been absorbed into the integral over the actual partition surface. This reproduces the Gioev-Klich-Widom formula, which is originally formulated for entanglement entropy but also applies to the scaling of charge fluctuation \cite{GioveKlich2006, Swingle2010, calabrese2012exact, Swingle2012, Song2011, Song2012, KunYang2012}. Next, we proceed to our main focus on the subleading $\log L$-terms. As we will see, the $\log L$-coefficient bears a similar mathematical structure as the Gioev-Klich-Widom formula in terms of a double-surface-integral over the Fermi surface in momentum space and the partition surface in real space.

\subsection{$\log L$-scaling and Fermi surface geometry}
Substituting in the subleading $-S^{(3)}(\hat{\bq})\abs{\bq}^3$ term in Eq.~\eqref{supp_eq: expansion of Sq} into Eq.~\eqref{supp_eq: starting point for charge fluctuation}, we obtain a logarithmic divergence from
\begin{equation}
    \int_0^{a^{-1}} d\abs{\tilde{\bq}}\mathcal{F}(\abs{\tilde{\bq}},L)^2\abs{\tilde{\bq}}^5 \supset 8\pi^2 \log(L/a),
\end{equation}
where $a$ is an ultra-violet cutoff at the lattice length scale and we have adopted (here and throughout) the symbol $\supset$ to indicate that the above expression on the right-hand-side is a particular subleading divergent term \textit{contained} in the full expression, so that we do not need to write out explicitly other terms with different types of divergence. Correspondingly, the charge fluctuation contains the following logarithmic divergence:
\begin{equation}
    \cc{Q^2_A} \supset -\left(\frac{1}{\pi}\int d^2\hat{\Omega}(\tilde{\bq})\;S^{(3)}(\hat{\bq})\frac{\abs{\bq}^3}{\abs{\tilde{\bq}}^3}\right)\log L \equiv -\Gamma\log L\;.
\end{equation}
We now perform the angular integral, in similar spirit to what we did in Eq.~\eqref{supp_eq: angular integral warmup}, with $S^{(3)}(\hat{\bq}) = S^{(3)}_{\rm G}(\hat{\bq})+S^{(3)}_{\rm QG}(\hat{\bq})$ and hence $\Gamma=\Gamma_{\rm G}+\Gamma_{\rm QG}$, using Eqs.~\eqref{supp_eq: gamma G} and ~\eqref{supp_eq: gamma QG}.

\subsubsection{Geometric contribution from $S^{(3)}_{\rm G}$}
From Eq.~\eqref{supp_eq: gamma G} we see that on each point of the Fermi surface (parametrized by $\bs$) we need to perform the following angular integral:
\begin{equation}
\begin{split}
    \int d^2\hat{\Omega}(\tilde{\bq})\;\delta(\hat{\bn}_F\cdot\hat{\bq}) I(\bs,\hat{\bq})\frac{\abs{\bq}^3}{\abs{\tilde{\bq}}^3} &= \int d^3\tilde{\bq}\;\delta(\abs{\tilde{\bq}}-1)\delta(\hat{\bn}_F\cdot\hat{\bq}) I(\bs,\hat{\bq}) \abs{\bq}^3\\
    &=\int d^3\bq\;\delta\left(\sqrt{q_a(R^{-1})^{ab}q_b}-1\right)\delta(\hat{\bn}_F\cdot\bq) I(\bs,\bq)\\
    &=\int d^3\bx \;\delta(\sqrt{x^aR_{ab}x^b}-1)\delta(\hat{\bn}_F\cdot R\bx) I(\bs, R\bx)\\
    &=\int'_{\partial A} dA_R \abs{\bn_R}^2 \delta(\hat{\bn}_F\cdot \hat{\bn}_R) I(\bs,\hat{\bn}_R)
\end{split}
\end{equation}
where the Fermi-surface geometric quantity $I(\bs,\hat{\bq})$ is defined in Eq.~\eqref{supp_eq: critical line integral and def of I}. The last line is an integral over the \textit{normalized} partition surface with $h(\bx)\equiv\sqrt{x^aR_{ab}x^b} =1$ (the actual partition surface of $A$ corresponds to $h(\bx)=L$), and $\hat{\bn}_R$ is the outward unit normal vector with $(\bn_R)_a \equiv \nabla_a h = R_{ab}x^b$. Consequently, the geometric contribution to the logarithmic divergence can be expressed $-\Gamma_{\rm G}\log L$ with 
\begin{equation}\label{supp_eq: Gamma G for general quadratic surface}
    \Gamma_{\rm G} =  \frac{1}{192\pi^4}\sum_m \int_{{\rm FS}_m}dA_F \int'_{\partial A}dA_R\abs{\bn_R}^2 \delta(\hat{\bn}_F\cdot \hat{\bn}_R) I(\bs,\hat{\bn}_R) = \int'_{\partial A} dA_R\;\abs{\bn_R}^2S^{(3)}_{\rm G}(\hat{\bn}_R)\;.
\end{equation}

\subsubsection{Quantum geometric contribution from $S^{(3)}_{\rm QG}$}
From Eq.~\eqref{supp_eq: gamma QG} we see that on each point of the Fermi surface (indexed by $m$) we need to perform the following angular integral:
\begin{equation}
\begin{split}
    \int d^2\hat{\Omega}(\tilde{\bq})\;\mathcal{G}^{ab}_m \hat{q}_a\hat{q}_b\abs{\hat{\bn}_F\cdot \hat{\bq}} \frac{\abs{\bq}^3}{\abs{\tilde{\bq}}^3}&=\int d^3\tilde{\bq} \;\delta(\abs{\tilde{\bq}}-1)\mathcal{G}^{ab}_m \hat{q}_a\hat{q}_b\abs{\hat{\bn}_F\cdot \hat{\bq}} \abs{\bq}^3\\
    &=\int d^3\bq\;\delta\left(\sqrt{q_a(R^{-1})^{ab}q_b}-1\right)\mathcal{G}^{ab}_m \hat{q}_a\hat{q}_b\abs{\hat{\bn}_F\cdot \hat{\bq}} \abs{\bq}^3\\
    &=\int d^3\bx \;\delta(\sqrt{x^aR_{ab}x^b}-1)\mathcal{G}^{ab}_m (R\bx)_a(R\bx)_b\abs{\hat{\bn}_F\cdot(R\bx)}\\
    &=\int'_{\partial A} dA_R\;\abs{\bn_R}^2 \abs{\hat{\bn}_F \cdot \hat{\bn}_R} \mathcal{G}^{ab}_m \hat{n}_{R,a} \hat{n}_{R,b}
\end{split}
\end{equation}
Again, the integral $\int'_{\partial A} dA_R$ refers to integrating over the normalized partition surface with $h(\bx)\equiv\sqrt{x^aR_{ab}x^b} =1$, $\bn_R \equiv \nabla h$, and the outward unit normal is denoted as $\hat{\bn}_R$. This leads to the quantum geometric contribution to the logarithmic divergence $-\Gamma_{\rm QG}\log L$ with
\begin{equation}\label{supp_eq: Gamma QG for general quadratic surface}
    \Gamma_{\rm QG} = \frac{1}{16\pi^4}\sum_m \int_{{\rm FS}_m} dA_F \int'_{\partial A}dA_R\;\abs{\bn_R}^2\abs{\hat{\bn}_F \cdot \hat{\bn}_R} \mathcal{G}^{ab}_m \hat{n}_{R,a} \hat{n}_{R,b}=\int'_{\partial A} dA_R\;\abs{\bn_R}^2S^{(3)}_{\rm QG}(\hat{\bn}_R)\;.
\end{equation}
This, together with Eq.~\eqref{supp_eq: Gamma G for general quadratic surface}, give Eq.~\eqref{eq: ellipsoid coefficient} in the main text.

\subsection{Spherical partition}
Next, we restrict to the spherical partition (i.e., region $A$ is a sphere and $R_{ab} = \delta_{ab}$ is the Euclidean metric) and further simplify Eqs.~\eqref{supp_eq: Gamma G for general quadratic surface} and ~\eqref{supp_eq: Gamma QG for general quadratic surface} into a more illuminating form. In this case, $\abs{\bn_R}=1$ and $\int'_{\partial A} dA_R$ is the integral over the unit sphere (or a solid angle integral). Let us invoke the following integral identities:
\begin{subequations}
\begin{align}
    \int_0^{2\pi}d\theta\;\prod_{i=1}^4(\hat{\bn}_R(\theta)\cdot\partial_{\mu_i}\bk_F) &= \frac{\pi}{4}(g_{\mu_1\mu_2}g_{\mu_3\mu_4}+g_{\mu_1\mu_3}g_{\mu_2\mu_4}+g_{\mu_1\mu_4}g_{\mu_2\mu_3})\;,\label{supp_eq: integral identity a} \\
    \int d^2\hat{\bn}_R\;\abs{\hat{\bn}_F\cdot\hat{\bn}_R}\hat{n}_{R,a}\hat{n}_{R,b} &= \frac{\pi}{2}(\delta_{ab} +\delta_{ac}\delta_{bd} \hat{n}_F^c \hat{n}_F^d)\;.\label{supp_eq: integral identity b} 
\end{align}
\end{subequations}
The first identity applies to the case where $\hat{\bn}_R$ and $\partial_{\mu_i}\bk_F$ all lie on the same 2D Euclidean plane. This is indeed the case for our interest because the delta function $\delta(\hat{\bn}_F\cdot \hat{\bn}_R)$ in Eq.~\eqref{supp_eq: Gamma G for general quadratic surface} constrains the $\hat{\bn}_R$-integral to the plane perpendicular to $\hat{\bn}_F$, which is also the plane where $\partial_\mu \bk_F$ lies (by definition of the normal vector: $\hat{n}_F^a\partial_\mu k_{Fa} = 0$). The metric $g_{\mu\nu}$ is the induced metric on the Fermi surface characterizing its shape, as defined in Eq.~\eqref{supp_eq: def of metric g}.\\

For $\Gamma_{\rm G}$, combining Eq.~\eqref{supp_eq: integral identity a} and Eq.~\eqref{supp_eq: Gamma G for general quadratic surface}, as well as the definition of $I(\bs,\hat{\bn}_R)$ in Eq.~\eqref{supp_eq: critical line integral and def of I}, we obtain
\begin{equation}
\begin{split}
    \int'_{\partial A}dA_R\abs{\bn_R}^2 \delta(\hat{\bn}_F\cdot \hat{\bn}_R) I(\bs,\hat{\bn}_R) &= \frac{\pi}{4} K_{\mu\nu} g^{\mu\sigma_1} g^{\nu\sigma_2} K_{\mu'\nu'} g^{\mu'\sigma_3} g^{\nu'\sigma_4} (g_{\sigma_1\sigma_2}g_{\sigma_3\sigma_4}+g_{\sigma_1\sigma_3}g_{\sigma_2\sigma_4}+g_{\sigma_1\sigma_4}g_{\sigma_2\sigma_3})\\
    &=\frac{\pi}{4}K_{\mu\nu}K_{\mu'\nu'}(g^{\mu\nu}g^{\mu'\nu'}+g^{\mu\mu'}g^{\nu\nu'}+g^{\mu\nu'}g^{\nu\mu'})\\
    &=\frac{\pi}{4}\left[(\tr K)^2+2\tr K^2\right] \\
    &=\frac{\pi}{4} [8\kappa_1\kappa_2+3(\kappa_1-\kappa_2)^2]
\end{split}
\end{equation}
where in the last line we have used Eq.~\eqref{supp_eq: relating K to curvature} to relate the second fundamental form $K_{\mu\nu}$ to the principal curvatures $(\kappa_1, \kappa_2)$ on the Fermi surface. This implies
\begin{equation}
    \Gamma_{\rm G} = \frac{1}{48\pi^2}\sum_m \left(\chi_m+\frac{3}{4\pi}\mathcal{W}_m\right) \geq \frac{1}{192\pi^2}\tilde{\chi}, \quad\quad \tilde{\chi}\equiv\sum_m(\chi_m+3\abs{\chi}_m)
\end{equation}
where $\chi_m = \frac{1}{2\pi}\int_{{\rm FS}_m}dA_F\; \kappa_1\kappa_2$ is the Euler characteristic of the $m$-th Fermi surface, and $\mathcal{W}_m = \frac{1}{4}\int_{{\rm FS}_m} dA_F (\kappa_1-\kappa_2)^2$ is the so-called Willmore energy for that Fermi surface which is non-negative measure of its anisotropy. The last inequality is obtained by applying the AM-GM inequality: $\kappa_1^2+\kappa_2^2\geq 2\abs{\kappa_1\kappa_2}$. Altogether, we obtain Eq.~\eqref{eq: Gamma_G} and Eq.~\eqref{eq: topological bound on Gamma_G} in the main text. \\

For $\Gamma_{\rm QG}$, combining Eq.~\eqref{supp_eq: integral identity b} and Eq.~\eqref{supp_eq: Gamma QG for general quadratic surface}, we obtain
\begin{equation}
\begin{split}
    \Gamma_{\rm QG}& = \frac{1}{32\pi^3}\sum_m \int_{{\rm FS}_m} dA_F\;\mathcal{G}^{ab}_m(\delta_{ab} +\delta_{ac}\delta_{bd} \hat{n}_F^c \hat{n}_F^d)\\
    &= \frac{1}{32\pi^3}\sum_m \int_{{\rm FS}_m} dA_F \left[\mathcal{G}_m^{\parallel_1 \parallel_1}+\mathcal{G}_m^{\parallel_2 \parallel_2}+2\mathcal{G}_m^{\perp\perp}\right] \geq \frac{1}{16\pi^2}\widetilde{C}, \quad\widetilde{C} \equiv \sum_m\abs{C_m}
\end{split}
\end{equation}
where in the last equality we have introduced a local orthogonal coordinate frame $\{\hat{\bee}^{\parallel_1},\hat{\bee}^{\parallel_2},\hat{\bee}^{\perp}\}$ at each point on the Fermi surface with $\hat{\bee}^{\parallel_{1,2}}$ tangential to the Fermi surface and $\hat{\bee}^{\perp}$ normal to it. Notice that the quantum metric obeys an inequality $\mathcal{G}^{\parallel_1 \parallel_1}+\mathcal{G}^{\parallel_2 \parallel_2} \geq 2\abs{\mathcal{F}^{\parallel_1\parallel_2}}$ \cite{Roy2014, inequality}, which implies $\int_{\rm FS} dA_F\; \mathcal{G}^{\parallel_1 \parallel_1}+\mathcal{G}^{\parallel_2 \parallel_2} \geq 2\abs{\int_{\rm FS} dA_F \mathcal{F}^{\parallel_1\parallel_2}}=2\pi\abs{C}$, where $C$ is the Fermi surface Chern number. Altogether we obtain Eq.~\eqref{eq: Gamma_Q} and Eq.~\eqref{eq: Gamma_Q bound} in the main text. 

\subsection{Ellipsoidal partition}

\subsubsection{Setup}
We now return to the general ellipsoidal partition surface defined in Eq.~\eqref{supp_eq: def of the real space partition surface}, and evaluate the generic logarithmic coefficient (c.f. Eq.~\eqref{eq: ellipsoid coefficient})
\begin{equation}
    \Gamma = \frac{1}{\pi}\int'_{\partial A} dA_R \abs{\bn_R}^2 S^{(3)}(\hat{\bn}_R)
\end{equation}
where the real-space integral is over the rescaled ellipsoidal surface $r^a R_{ab} r^b=1$ $(R=R^T>0,\;\det R=1)$.
To that end, it is useful to introduce $\eta=R^{1/2}$, which is the unique symmetric positive definite square root of $R$, and introduce a parametrization of the ellipsoid by a unit vector $\hat{\br}'$ as $\br=\eta^{-1}\hat{\br}'$. The outward normal is then $\bn_R= R\br=\eta\hat{\br}'$. With this change of variable,  it is easy to check that $dA_R  = \abs{\eta \hat{\br}'} d^2\hat{\br}'=\abs{\bn_R}d^2\hat{\br}'$ (see also the discussion around Eq.~\eqref{supp_eq: ellipsoid g and K}, but note that $\eta$ is defined differently). Thus
\begin{equation}
    \Gamma = \frac{1}{\pi}\int d^2\hat{\br}'S^{(3)}(\eta\hat{\br}') \implies \begin{cases}
        \Gamma_{\rm G} &= \frac{1}{192\pi^4}
    \int_{\rm FS} dA_F
    \int d^2\hat{\br}'\,
    \delta\!\left(
        \hat{\br}'\cdot\eta\hat{\bn}_F
    \right)
    \left[
        K^{\mu\nu}
        \bigl(\hat{\br}'\cdot\eta\partial_\mu\mathbf k_F\bigr)
        \bigl(\hat{\br}'\cdot\eta\partial_\nu\mathbf k_F\bigr)
    \right]^2 .\\
        \Gamma_{\rm QG} &= \frac{1}{16\pi^4}
    \int_{\rm FS}dA_F
    \int d^2\hat{\br}'\,
    |\hat{\br}'\cdot\eta\hat{\bn}_F|\,
    \mathcal G^{ab}_m
    (\eta\hat{\br}')_a(\eta\hat{\br}')_b .
    \end{cases}
\end{equation}
where for simplicity we use $\int_{\rm FS}$ to represent the entire $\sum_m \int_{{\rm FS}_m}$. The solid-angle integral $\int d^2\hat{\br}'$ can be easily performed similar to the integral identities in Eqs.~\eqref{supp_eq: integral identity a} and \eqref{supp_eq: integral identity b}, which gives us
\begin{equation}\label{supp_eq: Gamma G for ellipsoid 1}
    \Gamma_{\rm G} = \frac{1}{768\pi^3} \int_{\rm FS}dA_F \frac{1}{\abs{\eta\hat{\bn}_F}}\left[ (\gamma_{\mu\nu}K^{\mu\nu})^2+2\gamma_{\mu\sigma}\gamma_{\nu\tau}K^{\mu\nu}K^{\sigma\tau}\right],\quad \gamma_{\mu\nu}
    \equiv
    (\eta\partial_\mu\mathbf k_F)
    \cdot
    \mathbb{P}_{\eta\hat{\bn}_F}
    (\eta\partial_\nu\mathbf k_F)
\end{equation}
with $\mathbb{P}_{\eta\hat{\bn}_F} \equiv \mathds{1}-(\eta\hat{\bn}_F)(\eta\hat{\bn}_F)^T/\abs{\eta\hat{\bn}_F}^2$ the projector onto the plane orthogonal to $\eta\hat{\bn}_F$, and
\begin{equation}\label{supp_eq: Gamma QG for ellipsoid 1}
    \Gamma_{\rm QG} = 
    \frac{1}{32\pi^3}
    \int_{\rm FS}dA_F\,
    \abs{\eta \hat{\bn}_F}\,
    \mathcal G^{ab}_m
    \left[
        R_{ab}
        +
        \frac{
            (R\hat{\bn}_F)_a(R\hat{\bn}_F)_b
        }{
            \abs{\eta\hat{\bn}_F}^2
        }
    \right].
\end{equation}

\subsubsection{Linear deformation of the Fermi surface}
In order to cast $\Gamma_{\rm G}$ and $\Gamma_{\rm QG}$ into a more illuminating form (similar to the expressions for spherical partition), let us introduce the \textit{linearly deformed} Fermi surface, together with its induced metric tensor (induced from the standard Euclidean momentum space),  curvature tensor, and area element as
\begin{subequations}\label{supp_eq: linearly deformed FS}
\begin{align}
    {\rm FS}': \;\;\bk_F'(\bs) =\eta^{-1}\bk_F(\bs) , \;\;\partial_\mu\bk_F' = \eta^{-1}\partial_\mu
\bk_F,\;\;\hat{\bn}'_F = \frac{\eta\hat{\bn}_F}{\abs{\eta\hat{\bn}_F}},\;\;g'_{\mu\nu} = \partial_\mu\bk'_F\cdot\partial_\nu\bk'_F,\\
K'_{\mu\nu} = \frac{1}{\abs{\eta\hat{\bn}_F}}K_{\mu\nu},\quad\quad dA'_F = \hat{\bn}_F'\cdot(\partial_\mu\bk_F'\times \partial_\nu\bk_F') ds_1^\mu ds_2^\nu = \abs{\eta\hat{\bn}_F} dA_F
\end{align}
\end{subequations}
Notice that for this linearly deformed Femri surface (${\rm FS}'$), its tangent-plane projector is $\mathbb{P}_{\eta\hat{\bn}_F} = \partial_\mu \bk'_F (g')^{\mu\nu}(\partial_\nu \bk'_F)^T $, hence we have the following metric identity
\begin{equation}\label{supp_eq: metric identity for linear}
    \gamma_{\mu\nu} = g_{\mu\alpha} (g')^{\alpha\beta}g_{\beta\nu}
\end{equation}
In terms of quantum geometry, on ${\rm FS'}$ there is a corresponding quantum-state projector obtained as a ``pull-back" from the original Fermi surface, from which a quantum geometric tensor is introduced:
\begin{equation}
    {\rm FS}': \;\; P'(\bk_F') = P(\bk_F),\;\; \mathcal{Q'}^{ab}(\bk') \equiv \mathcal{G'}^{ab}-i\mathcal{F'}^{ab} = \Tr[P\frac{\partial P'}{\partial k'_a}\frac{\partial P'}{\partial k'_b}].
\end{equation}
In particular, we have $\partial'^aP' = \eta^a_b\partial^bP$, and thus
\begin{equation}\label{supp_eq: pullback quantum metric}
    \mathcal{G}'^{ab}(\bk'_F) = \eta^a_c \mathcal{G}^{cd}(\bk_F)\eta^b_d  \quad\text{and}\quad \mathcal{F}'^{ab}(\bk'_F) = \eta^a_c \mathcal{F}^{cd}(\bk_F)\eta^b_d 
\end{equation}

\subsubsection{Geometric contribution}
Combining Eqs.~\eqref{supp_eq: Gamma G for ellipsoid 1}, ~\eqref{supp_eq: linearly deformed FS} and~\eqref{supp_eq: metric identity for linear}, we arrive at an expression analogous to the case with spherical partition:
\begin{equation}\label{supp_eq: Gamma_G for ellipsoid using deformed FS}
\begin{split}
     \Gamma_{\rm G} &= \frac{1}{768\pi^3} \int_{\rm FS'}dA'_F \left[ (g'^{\mu\nu}K'_{\mu\nu})^2+2g'^{\mu\sigma}g'^{\nu\tau}K'_{\mu\nu}K'_{\sigma\tau}\right] = \frac{1}{768\pi^3} \int_{\rm FS'} dA'_F  [8\kappa'_1\kappa'_2 +3(\kappa'_1-\kappa'_2)^2]\\
     &= \frac{1}{48\pi^2} \sum_m \left(\chi'_m +\frac{3}{4\pi}\mathcal{W}_m'\right)  \geq \frac{1}{192\pi^2} \tilde{\chi}
\end{split}
\end{equation}
where $\kappa'_{1,2}$ are the principal curvatures of the linearly deformed Fermi surface, with $g'^{\mu\nu}K'_{\mu\nu}=\kappa'_1+\kappa'_2$, while $\chi'$ and $\mathcal{W}'$ are the corresponding Euler characteristic and Willmore energy of the deformed Fermi surface. In the last equality we have explicitly restored the summation over disconnected components of the Fermi surface. Notably, the \textbf{topological bound remains unchanged} when compared with the case of spherical partition,  as the linear deformation in Eq. ~\eqref{supp_eq: linearly deformed FS} leaves the Fermi surface topology unchanged. 

\subsubsection{Quantum geometric contribution}
Combining Eqs. ~\eqref{supp_eq: Gamma QG for ellipsoid 1}, ~\eqref{supp_eq: linearly deformed FS} and ~\eqref{supp_eq: pullback quantum metric}, we arrive at
\begin{equation}\label{supp_eq: Gamma_QG for ellipsoid using deformed FS}
\begin{split}
    \Gamma_{\rm QG}& = \frac{1}{32\pi^3}\sum_m \int_{{\rm FS'}_m} dA'_F\;\mathcal{G'}^{ab}_m(\delta_{ab} +\delta_{ac}\delta_{bd} \hat{n}_F^{'c} \hat{n}_F^{'d})\\
    &= \frac{1}{32\pi^3}\sum_m \int_{{\rm FS'}_m} dA'_F \left[\mathcal{G'}_m^{\parallel_1 \parallel_1}+\mathcal{G'}_m^{\parallel_2 \parallel_2}+2\mathcal{G'}_m^{\perp\perp}\right] \geq \frac{1}{16\pi^2}\widetilde{C}, \quad\widetilde{C} \equiv \sum_m\abs{C_m}
\end{split}
\end{equation}
which again formally resembles the expression for a spherical partition, but here in terms of the linearly deformed Fermi surface and its pull-back quantum metric. Notice that we \textbf{obtain the exact same Chern number bound} as in the case with spherical partition, which follows from the fact that $\mathcal{F}'\equiv \epsilon_{abc}\hat{n}_F^{'a}\mathcal{F'}^{bc} = \frac{1}{\abs{\eta\hat{\bn}_F}}\mathcal{F}$, and thus $\int_{\rm FS'} dA'_F \mathcal{F'} = \int_{\rm FS} dA_F \mathcal{F} = 2\pi C$, which is expected as the number of Berry monopoles (Weyl points) enclosed by the Fermi surface is invariant under any linear deformation. 

\subsubsection{Alternative interpretation}
There is an alternative interpretation of $\Gamma_{\rm G}$ and $\Gamma_{\rm QG}$ for a generic ellipsoidal partition surface, as derived above in Eqs.~\eqref{supp_eq: Gamma_G for ellipsoid using deformed FS} and ~\eqref{supp_eq: Gamma_QG for ellipsoid using deformed FS}, without referencing to the linearly deformed Fermi surface ($\rm{FS}'$) in Eq.~\eqref{supp_eq: linearly deformed FS}, but referencing the original Fermi surface ($\rm FS$). This is achieved by equipping the real space with a new metric ($h_{ab}=R_{ab}$), instead of the standard Euclidean metric ($h_{ab}=\delta_{ab}$) that we have been using so far. This way, the so-called ellipsoid in the standard Euclidean real space becomes a sphere under this new metric. Correspondingly, the metric in the reciprocal momentum space would be $h^{ab}=(R^{-1})^{ab}$, and the induced metric on the Fermi surface $\bk_F(\bs)$ would be $g^{(h)}_{\mu\nu}= \partial_\mu k_{F,a}h^{ab}\partial_\nu k_{F,b} = \partial_\mu \bk_F\cdot (R^{-1}\partial_\nu\bk_F)$, which equals to $g'_{\mu\nu}$ in Eq. ~\eqref{supp_eq: linearly deformed FS} as $R^{-1}=\eta^{-2}$. Notice that throughout this work $``\cdot"$ always represents the scalar product with respect to the standard Euclidean metric. Similarly, under the $h$-metric (with $h_{ab}=R_{ab}$), the area element, the unit normal and the curvature tensor can be expressed as follows:
\begin{equation}
\begin{split}
    g^{(h)}_{\mu\nu} &= g'_{\mu\nu}, \;\; dA_F^{(h)} = dA_F'=\abs{\eta\hat{\bn}_F} dA_F,\;\; \hat{\bn}^{(h)}_F = \frac{R\hat{\bn}_F}{\abs{\eta \hat{\bn}_F}},\\
    K^{(h)}_{\mu\nu}&=h(\hat{\bn}^{(h)}_F,\partial_\mu\partial_\nu\bk_F) = \hat{\bn}^{(h)}_F\cdot R^{-1}\partial_\mu\partial_\nu\bk_F= \frac{1}{\abs{\eta \hat{\bn}_F}}K_{\mu\nu}=K'_{\mu\nu}.
\end{split}
\end{equation}
where $h(\cdot, \cdot)$ is the scalar product with respect to the new metric $h^{ab}=(R^{-1})^{ab}$ in the momentum space. It is easy to check that $h(\hat{\bn}^{(h)}_F,\hat{\bn}^{(h)}_F)=1$ and $h(\hat{\bn}^{(h)}_F, \partial_\mu\bk_F)=0$, so indeed $\hat{\bn}^{(h)}_F$ is the unit normal to the Fermi surface under the $h$-metric.
From this perspective, we have
\begin{subequations}
    \begin{align}
        \Gamma_{\rm G}  &= \frac{1}{768\pi^3} \int_{\rm FS}dA^{(h)}_F \left[ (g^{(h)\mu\nu}K^{(h)}_{\mu\nu})^2+2g^{(h)\mu\sigma}g^{(h)\nu\tau}K^{(h)}_{\mu\nu}K^{(h)}_{\sigma\tau}\right] = \frac{1}{48\pi^2}\sum_m\left(\chi_m^{(h)}+\frac{3}{4\pi}\mathcal{W}^{(h)}_m\right) \\
        \Gamma_{\rm QG}&=\frac{1}{32\pi^3}
    \int_{\rm FS}dA^{(h)}_F\,
    \mathcal G^{ab}_m
    \left[
        R_{ab}
        + \hat{n}^{(h)}_{F,a} \hat{n}^{(h)}_{F,b}
    \right].
    \end{align}
\end{subequations}
which are analogous to Eqs.~\eqref{eq: Gamma_G} and~\eqref{eq: Gamma_Q}, with the only change being the change of metric. It is also clear from this perspective that topological bounds in Eqs.~\eqref{eq: topological bound on Gamma_G} and~\eqref{eq: Gamma_Q bound} remain unchanged. 

\section{Elementary derivation for $\Gamma_{\rm G}$ with spherical partition}\label{supp_sec: Gamma_G}
\setcounter{equation}{0}
\setcounter{figure}{0} 
Here we provide a derivation of Eq.~\eqref{eq: Gamma_G} directly from the elementary expression in Eq.~\eqref{eq: gamma_G 1}. Combining Eq.~\eqref{eq: def of Gamma} with Eq.~\eqref{eq: gamma_G 1}, we have
\begin{equation}
    \Gamma_{\rm G}=\frac{1}{\pi}\int d^2\hat{\bq}\;S^{(3)}_{\rm G}(\hat{\bq})=\frac{1}{192\pi^4}\int d^2\hat{\bq}\int dt\sum_p \frac{1}{\abs{r_p(t)}}
    \label{eq: cur-q}
\end{equation}
where $t$ is a coordinate in certain (arbitrary but fixed) direction orthogonal to $\bq$. The integral \eqref{eq: cur-q} can be shown to be a geometric integral over the Fermi surface (FS) as follows. We see that the integral obtains a contribution from all points of the FS, each point gives a contribution whenever $\hat{\bq}$ is parallel to the surface at the point. Let us parametrize the FS in the region of a point in the Monge representation as 
\begin{equation}
    z=\frac{x^2}{2R_x}+\frac{y^2}{2R_y},
    \label{eq: monge-rep}
\end{equation}
such that $\bq$ is parallel to the FS at $(0,0)$. We can write $\hat{\bq}=\hat{\bx}\cos\phi+\hat{\by}\sin{\phi}$, and choose the direction of integration $t$ to be orthogonal to $\hat{\bq}$:
\begin{equation}
    \hat{\bt}=\hat{\bx}\cos{\xi}\sin{\phi}-\hat{\by}\cos{\xi}\cos{\phi}+\hat{\bz}\sin\xi.
    \label{eq: s-dir-definition}
\end{equation}
Since $\xi$ was chosen arbitrarily, we expect its dependence to drop out. Given that choice, $r_p(t)$ is the radius of curvature in the direction perpendicular to both $\hat{\bm t},\hat{\bm q}$:
\begin{equation}
    \hat{\bq}\times\hat{\bt}= \hat{\bx}\sin{\xi}\sin{\phi}-\hat{\by}\sin{\xi}\cos{\phi}-\hat{\bz}\cos\xi.
\end{equation}
The radius $r(t)$ is then the radius of curvature of the FS in the plane perpendicular to $\bt$. We can parametrize this plane as $u\hat{\bq}+v\hat{\bq}\times\hat{\bt}$. Substituting in \eqref{eq: monge-rep} gives
\begin{equation}
    -v\cos\xi=\frac{(v \sin\xi\sin\phi+u\cos\phi)^2}{2R_x}+\frac{(-v\sin\xi\cos\phi+u\sin\phi)^2}{2R_y}. 
\end{equation}
The radius of curvature is then obtained as
\begin{equation}
    \frac{1}{\abs{r(t)}}=\abs{\partial_u^2v|_{u=v=0}}=\frac{1}{\abs{\cos\xi}}\abs{\frac{\cos^2\phi}{R_x}+\frac{\sin^2\phi}{R_y}}.
    \label{eq: rs-result}
\end{equation}
% From \eqref{eq: rs-result} we see that we can parametrize the path by arc length $dl$ projected on the plane perpendicular to $\bq$. 
When $\bq$ deviates from the $x,y$ plane, it can be written as $\bq=\hat{\bx}\cos\phi+\hat{\by}\sin\phi+\theta\hat{\bz}$. We want to perform a coordinate transformation from $\theta,\phi,t$ to $x,y,\phi$. Integrating over $\phi$ then gives a Fermi surface integral. We have
\begin{equation}
\begin{aligned}
    d\theta dt d\phi = J dxdyd\phi, \quad\quad
    J= \abs{\begin{array}{cc}
        \partial_x\theta & \partial_y \theta \\
        \partial_x t & \partial_y t 
    \end{array}}.
\end{aligned}
\end{equation}
The relation between $x,y$ and $\theta$ is obtained by the requirement that $\bq$ has to be parallel to the FS, hence
\begin{equation}
    \theta=\frac{x\cos\phi}{R_x}+\frac{y\sin\phi}{R_y}. 
\end{equation}
For $t$, we have
\begin{equation}
    t=(x\hat{\bx}+y\hat{\by})\cdot\hat{\bt}=\cos\xi(x\sin{\phi}-y\cos\phi).
\end{equation}
The Jacobian is therefore
\begin{equation}
    J=\abs{\cos\xi\left(\frac{\cos^2\phi}{R_x}+\frac{\sin^2\phi}{R_y}\right)}. 
\end{equation}
Putting all these together, we obtain
\begin{equation}
    \Gamma_{\rm G}=\frac{1}{192\pi^4}\int_{\rm FS}dxdy \int d\phi \left(\frac{\cos^2\phi}{R_x}+\frac{\sin^2\phi}{R_y}\right)^2=\frac{1}{768\pi^3}\int_{\rm FS}  8\kappa_x\kappa_y+3(\kappa_x-\kappa_y)^2 dxdy
\end{equation}
where we have introduced the principal curvatures $\kappa_{x,y}=1/R_{x,y}$. The integral of the first term can be identified as the Euler characteristic of the Fermi surface, by the Gauss-Bonnet theorem and is a topological invariant, while the second term gives the so-called Willmore energy,
\begin{equation}
    \chi_{\mathrm{FS}}= \frac{1}{2\pi}\int_{\rm FS} \kappa_x \kappa_y dx dy,\quad\mathcal{W}_{\rm FS} = \frac{1}{4}\int_{\rm FS} (\kappa_x-\kappa_y)^2 dxdy. 
\end{equation}
The Willmore energy is a non-negative measure of anisotropy of the surface deviating from a sphere. It is not a topological invariant, but is, interestingly an invariant under conformal transformations \cite{white1973global}.

\section{Structure factor in a Weyl metal}\label{supp_sec: Weyl}
\setcounter{equation}{0}
\setcounter{figure}{0}

Consider the equal-time fermion Green function for the Weyl Hamiltonian $H(\bk) = \bk\cdot\boldsymbol{\sigma}$:
\begin{equation}
\begin{split}
    G_{\alpha\beta}(\bR-\bR') &\equiv \frac{i}{V_c}\langle c^\dagger_{\bR' \beta} c_{\bR\alpha}\rangle = \frac{i}{V} \sum_{\bk, m} e^{i\bk\cdot(\bR-\bR')}\theta(E_F-E_{m,\bk}) [P_m (\bk)]_{\alpha\beta} \\
    &= i\sum_{m=\pm} \int \frac{d^3\bk}{(2\pi)^3} e^{i\bk\cdot(\bR-\bR')}\theta(E_F-m\abs{\bk}) \frac{1}{2}\left(\mathds{1}+m \hat{\bk}\cdot\boldsymbol{\sigma}\right)_{\alpha\beta}
\end{split}
\end{equation}
where $\alpha,\beta$ are orbital (pseudo-spin) indices, $V_c$ is the unit-cell volume and $V$ is the total system volume. We take $\bR_\alpha=\bR$ (so all intra-cell orbitals coincide) and consider the continuum limit ($V_c\rightarrow0$, $\bR$ is continuously valued) and the thermodynamic limit ($V\rightarrow \infty$).
Assume $E_F<0$, so only the lower band $m=-1$ is filled with a Fermi surface at $k_F$, then
\begin{equation}
    G(\br) = \frac{i}{2} \int \frac{d^3\bk}{(2\pi)^3} e^{i\bk\cdot\br}\theta(\abs{\bk}-k_F)\left(\mathds{1}- \frac{\bk}{\abs{\bk}}\cdot\boldsymbol{\sigma}\right).
\end{equation}
We now introduce two scalar functions
\begin{flalign}
f_{1}(r) =\int\frac{d^3\bk}{(2\pi)^{3}}e^{i\bk\cdot\br}\theta(|\bk|-k_{F})=\frac{k_{F}r\cos(k_{F}r)-\sin(k_{F}r)}{2\pi^{2}r^{3}},\;\;
f_{2}(r)  =\int\frac{d^3\bk}{(2\pi)^{3}}e^{i\bk\cdot\br}\frac{\theta(|\bk|-k_{F})}{|\bk|}=\frac{\cos(k_{F}r)}{2\pi^{2}r^{2}}.
\end{flalign}
Using these functions, we can write the equal-time Green function
as $G(\br)=\frac{1}{2}\left[f_{1}(r)\mathds{1}+if_{2}^{\prime}(r)\hat{\br}\cdot\boldsymbol{\sigma}\right]$. The real-space equal-time density-density correlation ($S(\br) \equiv \cc{\rho(\br)\rho(0)}$) is then 
\begin{equation}\label{supp_eq: Sr}
    \begin{split}
        S(\br) = \sum_{\alpha,\beta} G_{\alpha\beta}(\br) G_{\beta\alpha}(-\br)= -\frac{1}{4}\Tr[\left(f_1(r)\mathds{1}+if'_2(r)\hat{\br}\cdot\boldsymbol{\sigma}\right)^2]=-\frac{1}{16\pi^{4}r^{6}}[2k_{F}^{2}r^{2}+2k_{F}r\sin(2k_{F}r)+3\cos(2k_{F}r)+5].
    \end{split}
\end{equation}
The Fourier transform of the above expression (upon using dimensional regularization or imposing a short-distance cutoff on the integral) gives
\begin{equation}
    S(\bq) = \int d^3\br\; e^{i\bq\cdot\br}S(\br) = \frac{k_F^2}{8\pi^2}\abs{\bq} - \frac{5}{192\pi^2}\abs{\bq}^3 + ...
\end{equation}
for $\abs{\bq}<2k_F$, where the ellipsis represent analytic terms that depend on the short-distance cutoff or regularization. Hence $S^{(3)}(\hat{\bq}) = 5/(192\pi^2)$ for a Weyl metal with a finite Fermi surface, as discussed in the main text (Example 2). Notice that the $\cos(2k_F r)$ term inside the square bracket of Eq.~\eqref{supp_eq: Sr}, related to the Friedel oscillation, does not contribute to the $\abs{\bq}^3$ term due to a cancellation from the oscillation.  On the other hand, if $E_F=0$ ($k_F=0$) where the Fermi surface has diminished into a point ($k_F=0$), the constant term inside the square bracket of Eq.~\eqref{supp_eq: Sr} becomes 8 instead of 5, and subsequently for a Weyl semimetal, $S^{(3)}(\hat{\bq})=1/(24\pi^2)$, which reproduces the known CFT result \cite{Mora2019, Wu_fluc_FS}.

\section{Structure Factor in Landau Fermi Liquids}\label{supp_sec: FL}
\setcounter{equation}{0}
\setcounter{figure}{0} 
The long-wavelength behavior of the static structure factor in a Fermi liquid is qualitatively identical to that of free fermions: only odd powers of $|\bq|$ appear in $S(\bq)$, with local interactions modifying their coefficients. By either solving the Landau kinetic equation or performing a random phase approximation (RPA) analysis of the density-density correlator, the Matsubara density response is given by~\cite{Giuliani_Vignale_2005}
\begin{align}
\Pi_{\tau\tau}^{\textrm{RPA}}(i\omega,\bq)=\frac{\Pi_{\tau\tau}(i\omega,\bq)}{1-(F_{0}/\mathscr{D}_{F})\Pi_{\tau\tau}(i\omega,\bq)},
\label{eq:_FL_RPA}
\end{align}
where $F_{0}$ is the (dimensionless) Landau parameter in the $s$-wave channel, and $\mathscr{D}_{F}=mk_F/(2\pi^2)$ is the density of states at the Fermi surface in three spatial dimensions. The structure factor is $S(\bq)= -\int_{-\infty}^{+\infty}\frac{d\omega}{2\pi} \Pi_{\tau\tau}^{\rm RPA} (i\omega,\bq)$.

Introducing the dimensionless variables $\tilde{\omega}=\frac{m\omega}{k_{F}|\bq|}$ and $\tilde{k}=\frac{|\bq|}{2k_{F}}$, the Lindhard function takes the form~\cite{Giuliani_Vignale_2005}
\begin{equation}
\Pi_{\tau\tau}(i\omega,\bq)=\mathscr{D}_{F}\frac{\Psi_3(i\tilde{\omega}-\tilde{k})-\Psi_3(i\tilde{\omega}+\tilde{k})}{2\tilde{k}}\quad\textrm{with}\quad \Psi_{3}(z)= \frac{z}{2}+\frac{1-z^2}{4}\ln\frac{z+1}{z-1}.
\end{equation}
For $|\bq|<2k_{F}$ (i.e., $\tilde{k}<1$), the Lindhard function admits the Taylor series
\begin{align}
\Pi_{\tau\tau}(i\omega,\bq)=-\mathscr{D}_{F}\sum_{\ell=0}^{+\infty}\frac{\Psi_{3}^{(2\ell+1)}(i\tilde{\omega})}{(2\ell+1)!}\tilde{k}^{2\ell},
\end{align}
where $\Psi_{3}^{(n)}$ denotes the $n$-th derivative of $\Psi_{3}$. Substituting the above into Eq.~\eqref{eq:_FL_RPA}, the structure factor acquires the following long-wavelength expansion (normalized by the charge density $\langle\rho \rangle = k_F^3/(6\pi^2)$)
\begin{equation}
\begin{split}
    S(\bq)/\langle\rho\rangle = - 6\tilde{k} \int_{-\infty}^{+\infty}\frac{d\tilde{\omega}}{2\pi}\frac{\Pi_{\tau\tau}(i\omega,\bq)/\mathscr{D}_{F}}{1-F_{0}\Pi_{\tau\tau}(i\omega,\bq)/\mathscr{D}_{F}} = \sum_{\ell=0}^{+\infty}\int_{-\infty}^{+\infty}\frac{d\tilde{\omega}}{2\pi}g_{2\ell+1}(i\tilde{\omega})\tilde{k}^{2\ell+1},
\end{split}
\end{equation}
where, for example,
\begin{align}
g_{1}(i\tilde{\omega})=\frac{6\Psi_{3}^{(1)}(i\tilde{\omega})}{1+F_{0}\Psi_{3}^{(1)}(i\tilde{\omega})},\qquad g_{3}(i\tilde{\omega})=\frac{\Psi_3^{(3)}(i\tilde{\omega})}{(1+F_{0}\Psi_{3}^{(1)}(i\tilde{\omega}))^{2}},
\end{align}
and similarly for higher orders. This makes manifest that only odd powers of $|\bq|$ appear in $S(\bq)$, in agreement with the structure found in the free-fermion case. In terms of the expansion coefficient introduced in Eq.~\eqref{eq: expansion of Sq}, we have
\begin{equation}
    S^{(1)}(\hat{\bq}) = \frac{k_F^2}{24\pi^3}\int_{-\infty}^{+\infty} d\omega\; g_1(i\omega)\quad\text{and}\quad S^{(3)}(\hat{\bq}) = -\frac{1}{96\pi^3}\int_{-\infty}^{+\infty} d\omega\; g_3(i\omega).
\end{equation}
Their dependence on the Landau parameter $F_0$ is shown in Fig.~\ref{fig: RPA} of the main text. While our analysis here focuses on spatial dimension $D=3$, it can be straightforwardly generalized to other dimensions. For $D=2$, the dependence of the linear $|\bq|$ term (i.e., $S^{(1)}$) on $F_{0}$ has been analyzed in Refs.~\cite{Wu_fluc_FS, SwingleSenthil2013, cai2024disorder}.

\end{document}